\documentclass[12pt]{article}  
\usepackage{graphicx}
\usepackage{epsfig}
\usepackage{hyperref}
\usepackage {amsmath}
\usepackage {amssymb}
\usepackage {bm}  
\usepackage[usenames,dvipsnames]{color}
\usepackage{xspace} 
\newcommand{\pt}{\mbox{$p_T$}\xspace}

\newcommand{\Npart}{\mbox{$N_{\rm part}$}\xspace}
\newcommand{\Ncoll}{\mbox{$N_{\rm coll}$}\xspace}
\newcommand{\Nch}{\mbox{$N_{\rm ch}$}\xspace}
\newcommand{\Et}{\mbox{${\rm E}_T$}\xspace}

\newcommand{\sqs}{\mbox{$\sqrt{s}$}\xspace}
\newcommand{\sqsn}{\mbox{$\sqrt{s_{_{NN}}}$}\xspace}

\newcommand{\Nqp}{\mbox{$N_{qp}$}\xspace}
\def\lsim{\raise0.3ex\hbox{$<$\kern-0.75em\raise-1.1ex\hbox{$\sim$}}}
\def\gsim{\raise0.3ex\hbox{$>$\kern-0.75em\raise-1.1ex\hbox{$\sim$}}}

\def\mean#1{\left<#1\right>}
\def\Journal#1#2#3#4{{#1}{\bf #2} (#4) #3}

\def\NIMA{{Nucl. Instrum. Methods A}}
\def\NPA{{Nucl. Phys. A}}
\def\NPB{{Nucl. Phys. B}}
\def\PLB{{Phys. Lett. B}}

\def\PL{Phys. Lett.\ }
\def\PRL{Phys. Rev. Lett.\ }
\def\PRD{{Phys. Rev. D}}
\def\PRC{{Phys. Rev. C}}

\def\RMP{Rev. Mod. Phys.\ }

\def\QGP{{\color{Red} Q}{\color{Blue} G}{\color{Green} P}} 
\def\QCD{{\color{Red} Q}{\color{Green} C}{\color{Blue} D}}

\normalsize
\begin{document}

\title{Highlights from BNL and RHIC 2016}
\author{M.~J.~Tannenbaum
\thanks{Research supported by U.~S.~Department of Energy, DE-SC0012704.}
\\Physics Department, 510c,\\
Brookhaven National Laboratory,\\
Upton, NY 11973-5000, USA\\
mjt@bnl.gov} 
\date{}
\maketitle
\vspace*{-2pc}
\section{Introduction}\label{sec:introduction}
\begin{figure}[!h]
\begin{center}
\includegraphics[width=0.95\textwidth]{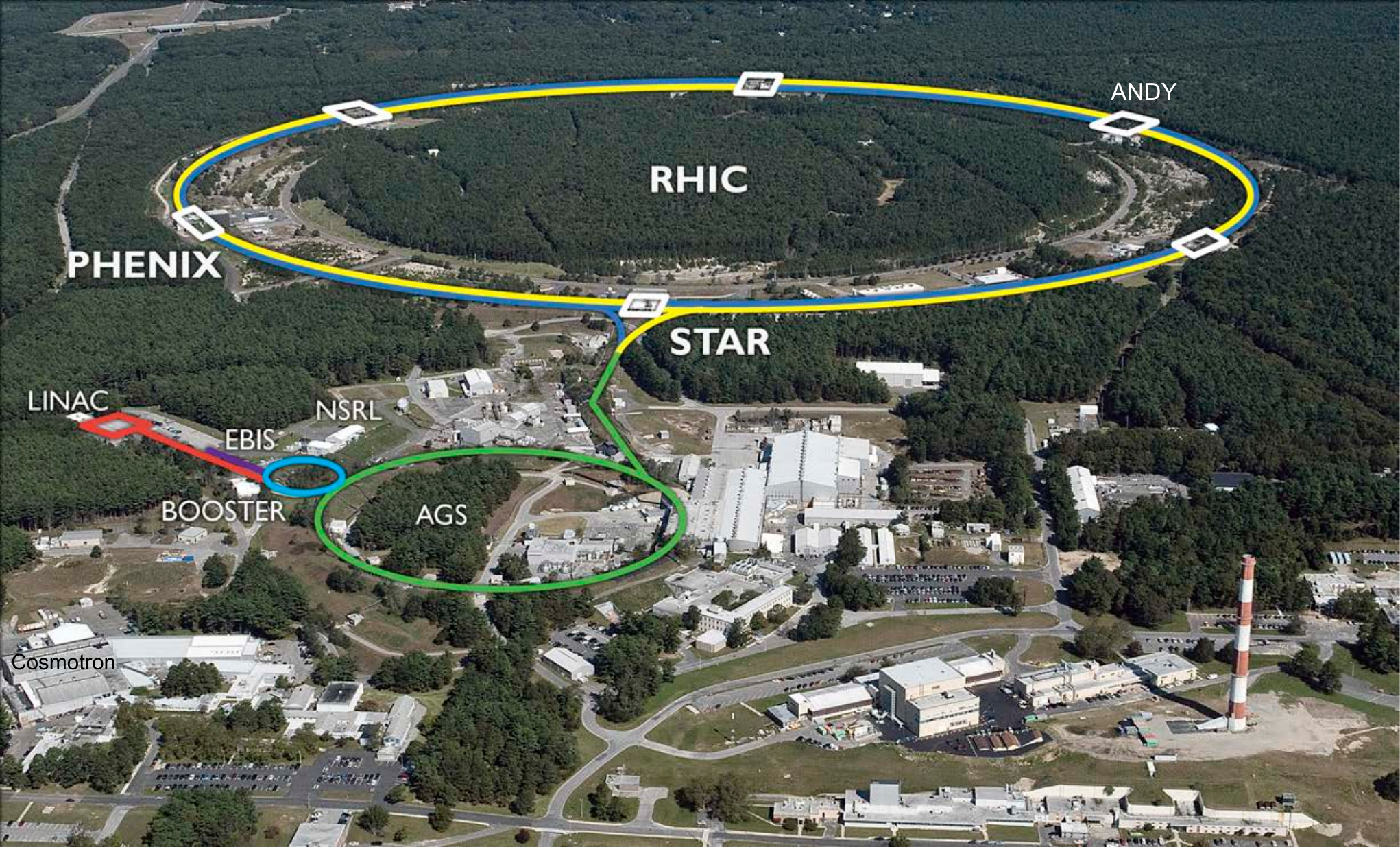}
\end{center}\vspace*{-1.5pc}
\caption[]{\footnotesize Aerial view of BNL with Cosmotron, AGS, RHIC with Injection line and Experiments indicated.}
\label{fig:BNLair}\vspace*{-0.5pc}
\end{figure}

The Relativistic Heavy Ion Collider (RHIC) was designed and built at Brookhaven National Laboratory (BNL)~\cite{NIMA499} and is one of the two remaining operating hadron colliders in the world, the other being the CERN LHC. There were also several previous facilities at BNL (Fig.~\ref{fig:BNLair}) including the 30 GeV Alternating Gradient Synchrotron (AGS--now the RHIC injector) where the muon-neutrino~\cite{Danby1962}, and CP violation~\cite{CPAGS} were discovered and the 3 GeV Cosmotron where the $K^0_2$ was discovered~\cite{LMLKL}. There have been many other discoveries at BNL~\cite{egseeMJT2015}. 

\begin{figure}[!htb] 
	\begin{center}
\includegraphics[width=0.44\textwidth]{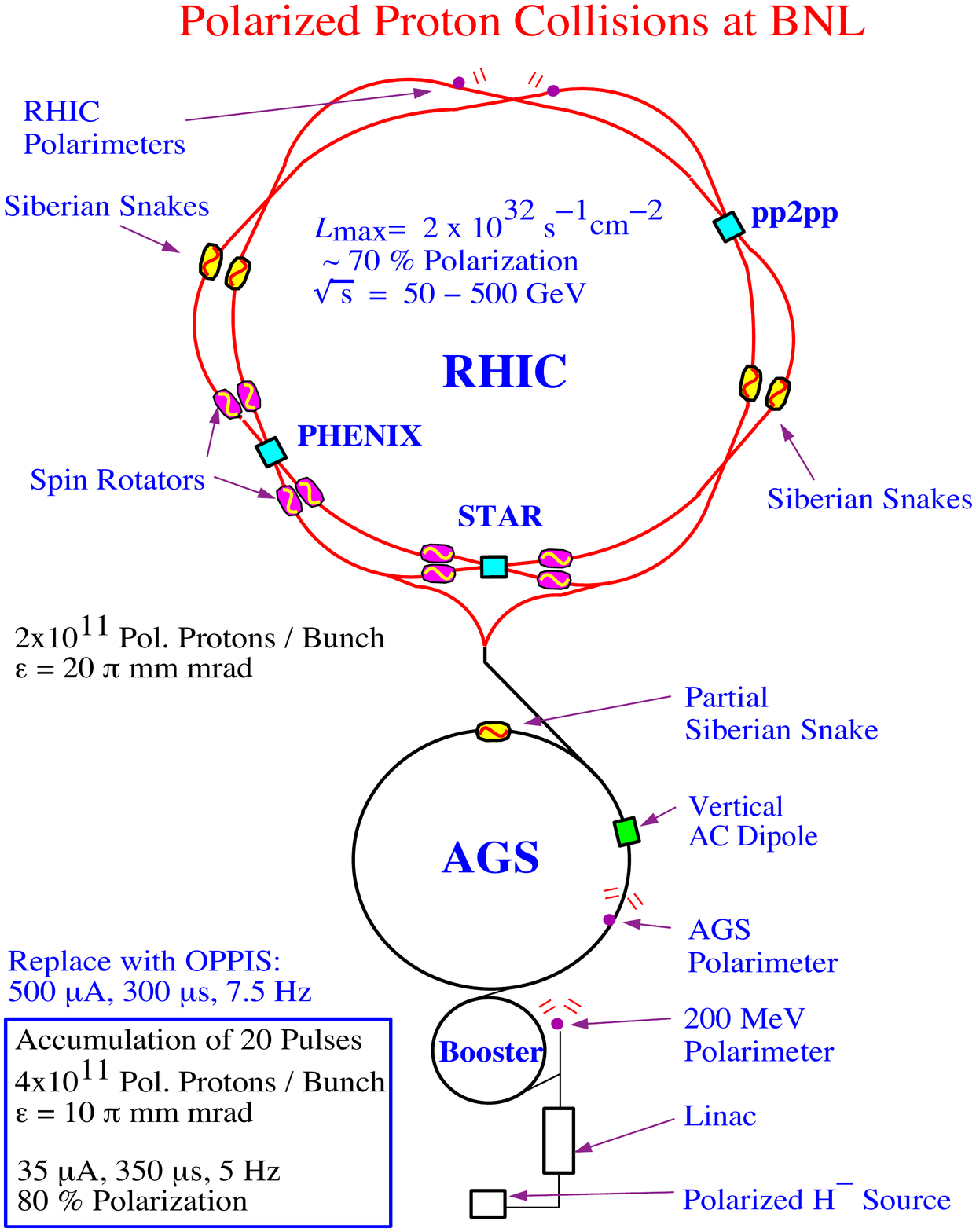}\hspace*{1pc}
\includegraphics[width=0.42\textwidth]{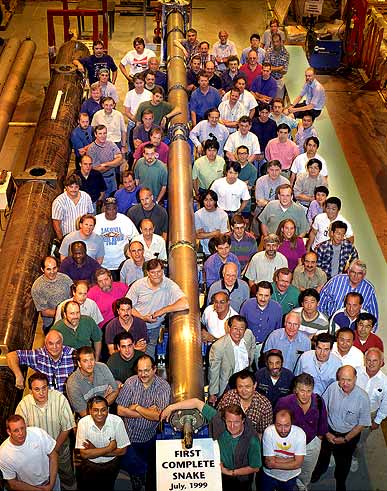}
\end{center}\vspace*{-1.5pc}
\caption[]{\footnotesize a) (Left) RHIC machine with polarized proton hardware highlighted. The spin rotators enable the proton spins to be rotated from their equilibrium transverse polarization to longitudinal at the experiments while the siberian snakes flip the spin direction half way around the ring to preserve the polarization by canceling imperfections. b)(right) Completion of first siberian snake magnet wound at BNL.    
\label{fig:RHICmachine}}
\end{figure}

In addition to being able to accelerate and collide any nucleus with any other nucleus, e.g. Cu$+$Au, RHIC is also the first and only polarized proton collider (Fig.~\ref{fig:RHICmachine}). 

	\vspace{+1.5pc}
\section{Experiments and Detectors at RHIC}
\begin{figure}[!tb] 
	\begin{center}
\raisebox{0pc}{\includegraphics[width=0.48\textwidth]{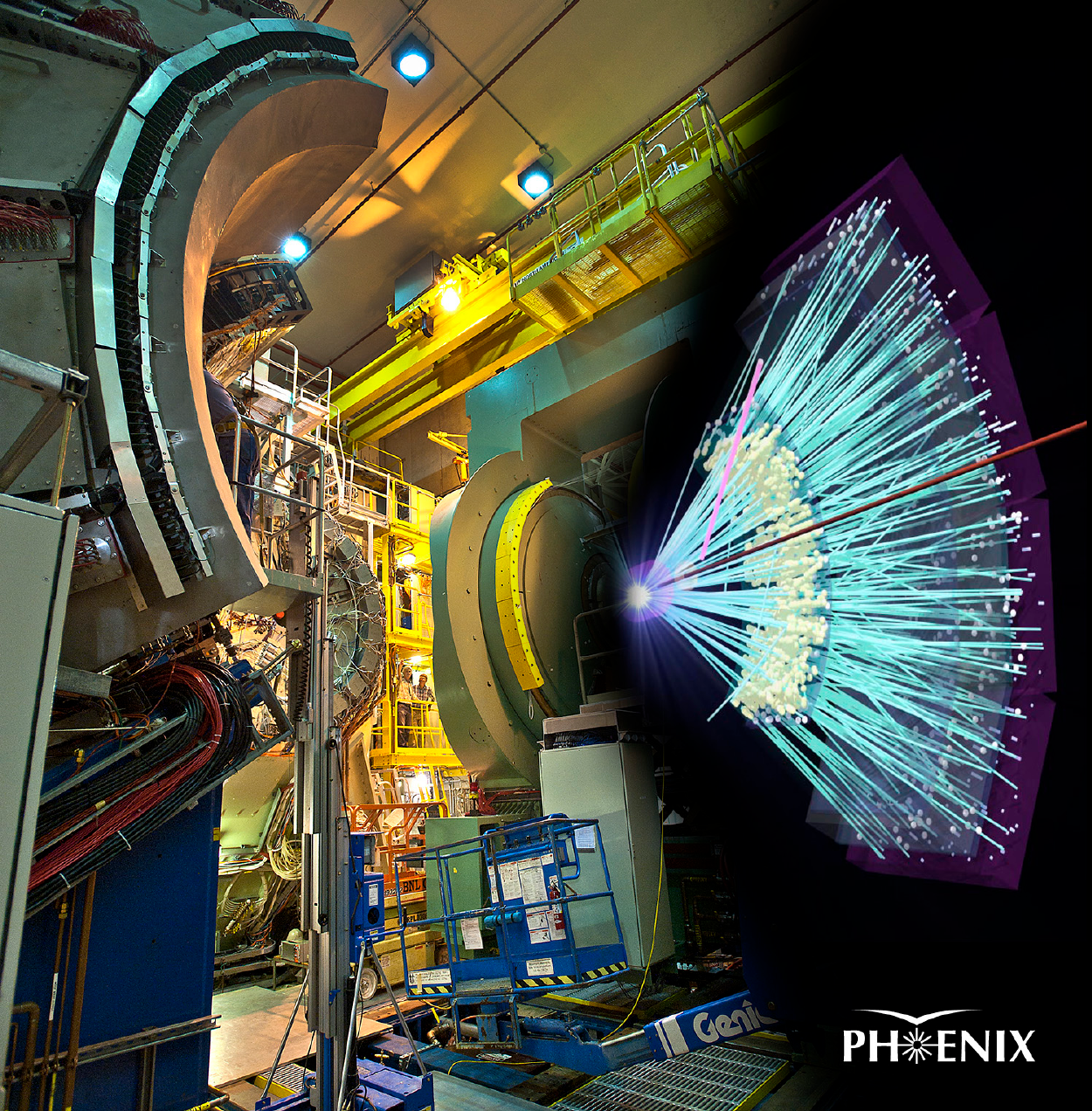}}\hspace*{1pc}
\raisebox{0pc}{\includegraphics[width=0.48\textwidth]{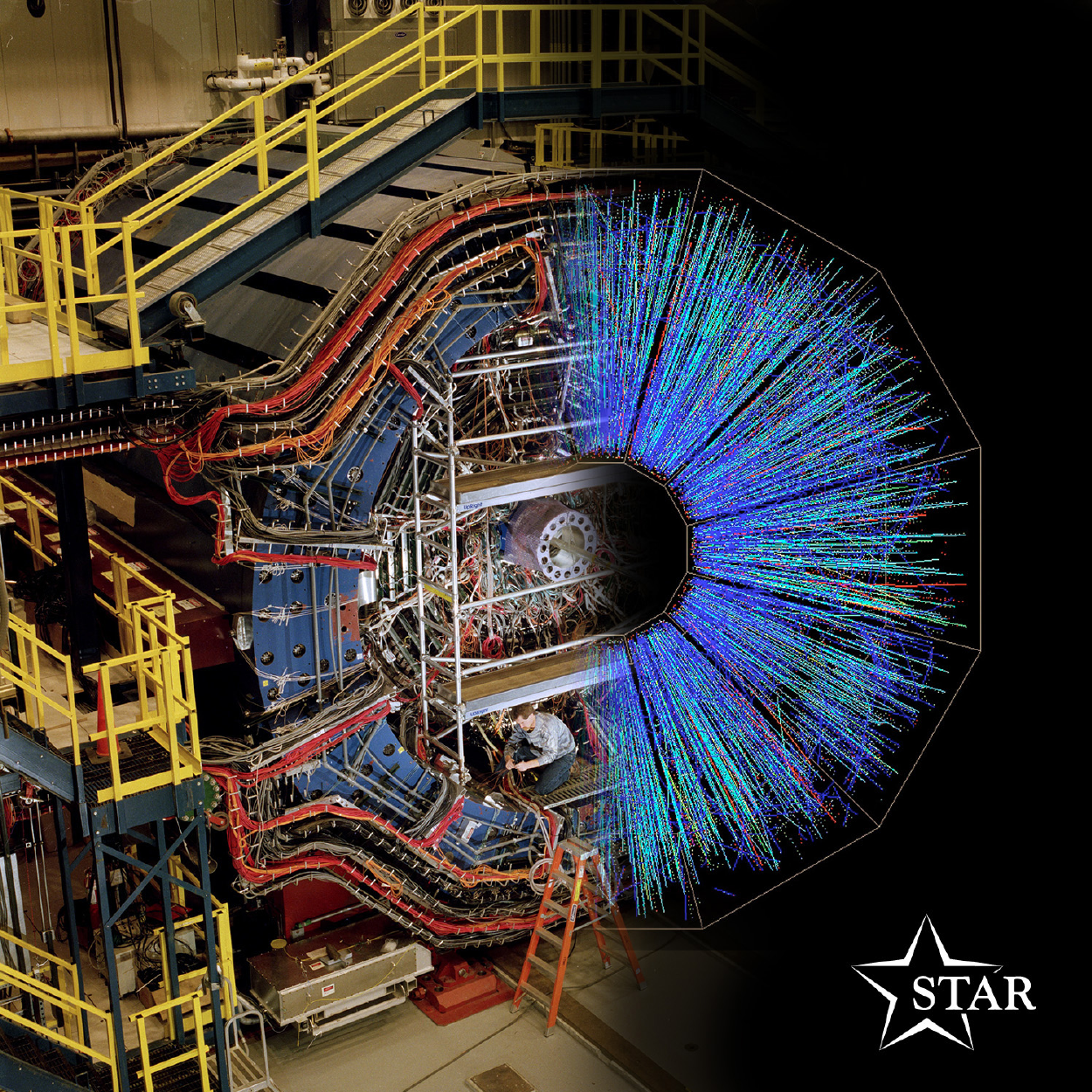}}
\end{center}\vspace*{-1.5pc}
\caption[]{\footnotesize a) (left) PHENIX detector with central spectrometer moved out of position and central magnet visible b)(right) STAR detector. Both detectors have event displays superimposed.     
\label{fig:PXSTAR}}
\end{figure}

    Figure~\ref{fig:BNLair} now shows three experiments: PHENIX and STAR, the two major detectors that have been operating with several upgrades since the start of operations in the year 2000, along with a new experiment ANDY, which is a special-purpose forward EM and Hadron Calorimeter detector to measure the transverse single-spin asymmetry (AN) of Drell-Yan (DY) pairs in p$\uparrow +$p collisions. PHENIX is a two-arm spectrometer with a fine grain EM calorimeter, Ring Imaging Cerenkov counter, time-of-flight (TOF) and drift-chamber tracking for $e^{\pm}$, $\gamma$ and identified hadron measurements at mid-rapidity, with muon spectrometers at forward and backward rapidity; while STAR is a more conventional solenoid with full azimuthal coverage, a TPC tracker, a Barrel EM calorimeter inside the magnet coil and TOF for particle identification. Both experiments have micro-vertex detectors (Fig.~\ref{fig:PXSTAR}).
\subsection{ANDY and other transverse single spin asymmetry measurements}
\begin{figure}[!h]
\begin{center}     
\raisebox{1pc}{\includegraphics[width=0.50\textwidth]{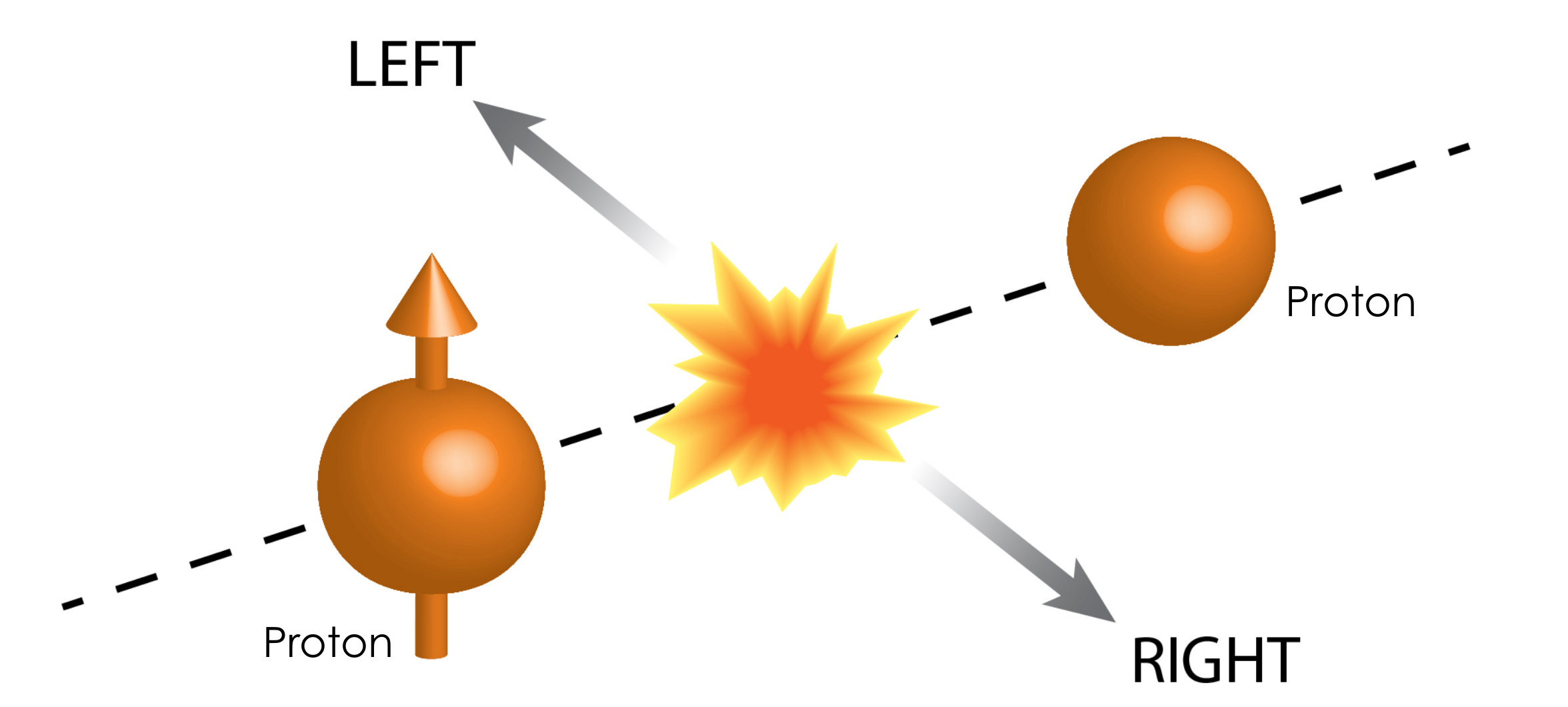}}
\raisebox{0pc}{\includegraphics[width=0.48\textwidth]{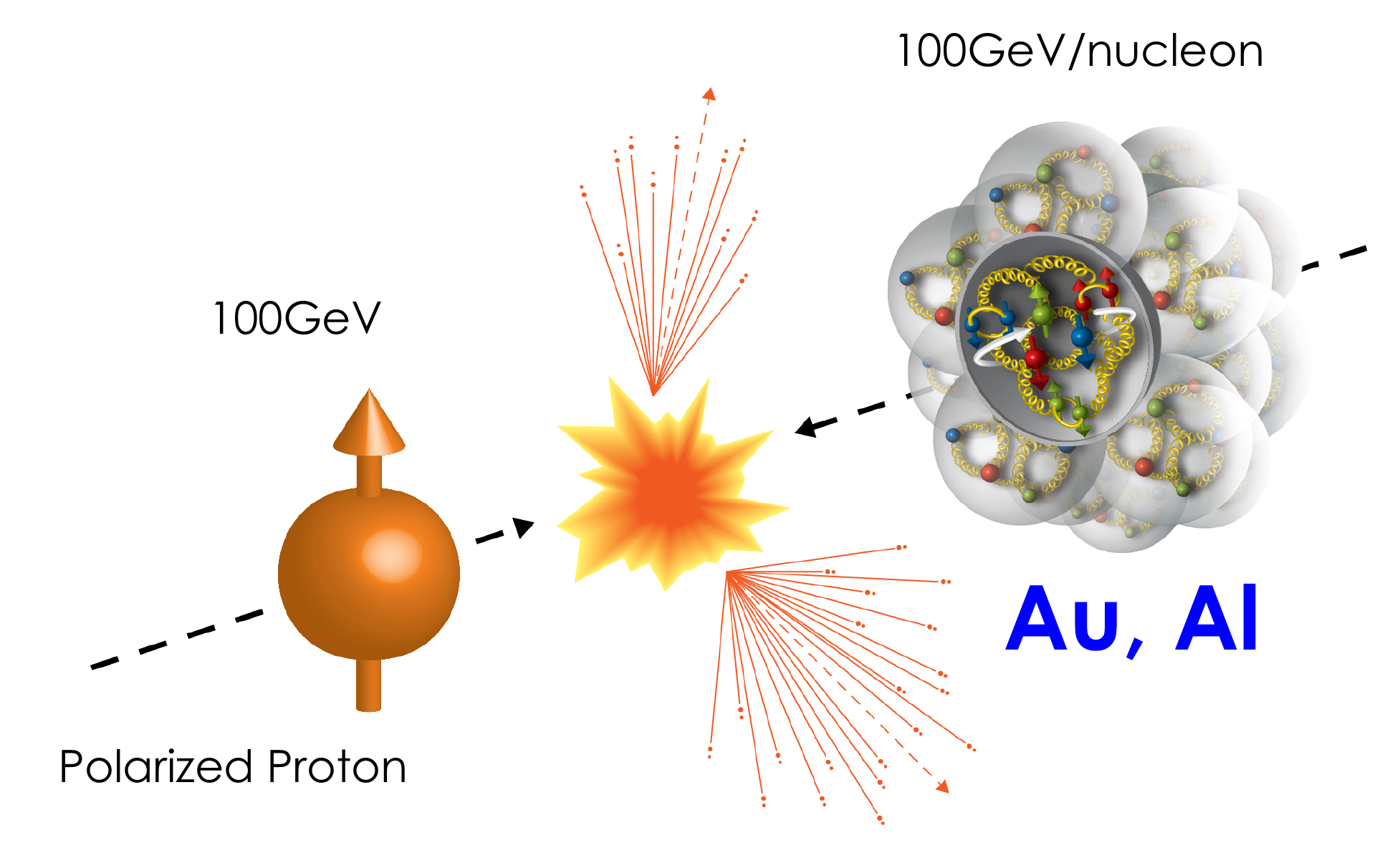}}
\end{center}\vspace*{-1.5pc}
\caption[]{\footnotesize a) p$\uparrow +$p scattering with L, R measurement. b) p$\uparrow +$A scattering~\cite{ItaruMENU}.     
\label{fig:AN}}
\end{figure}

The transverse single spin asymmetry ($A_N$) is basically a left-right asymmetry either in elastic scattering of a polarized proton on a target or in production of an identified particle or a Drell-Yan $e^+ e^-$ or $\mu^+ \mu^-$ pair (Fig~\ref{fig:AN}). It is usually measured by flipping the proton spin---comparing counts with the proton spin up versus down. \pagebreak

\begin{figure}[!h]
\begin{center}     
\raisebox{0pc}{\includegraphics[width=0.58\textwidth]{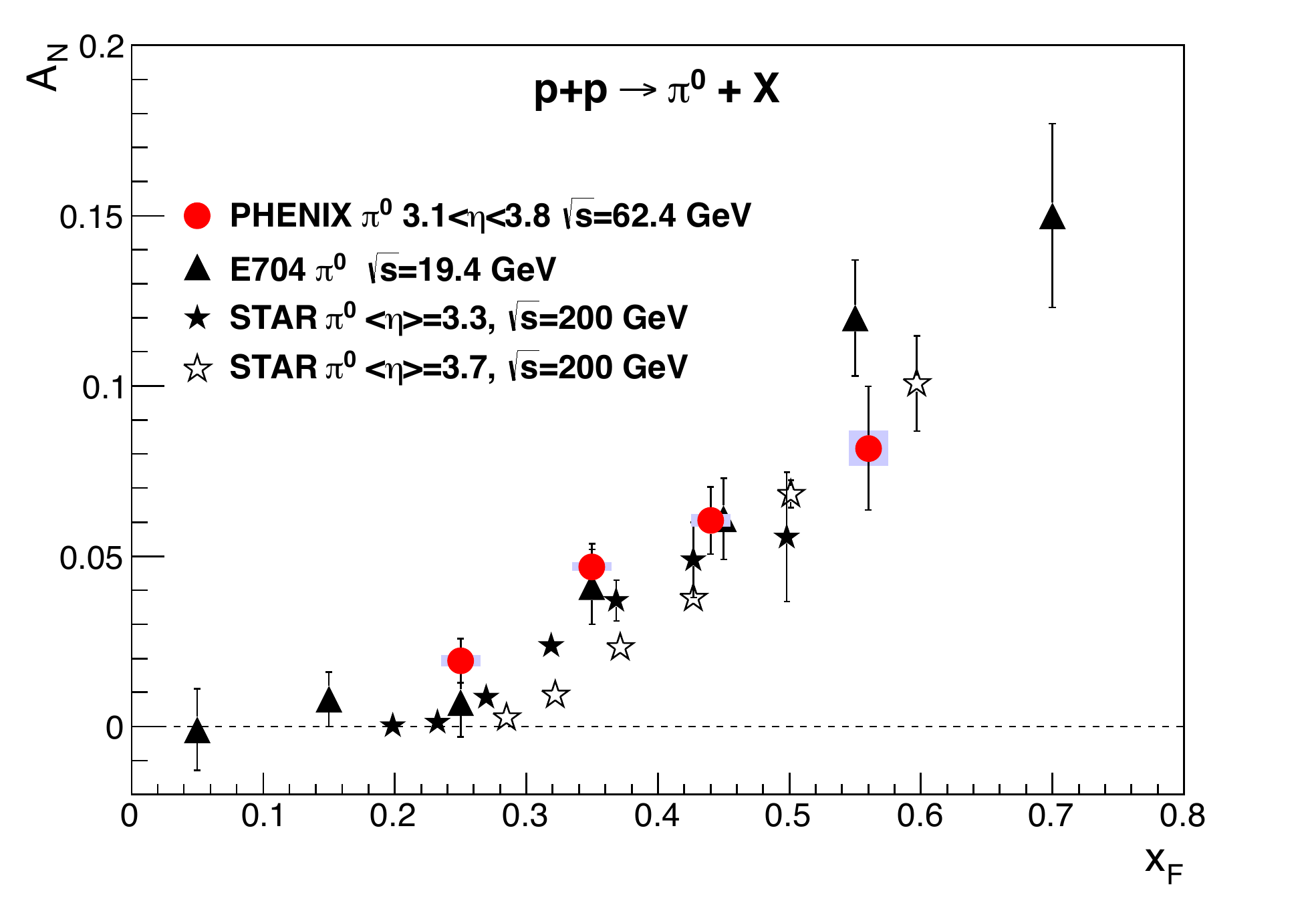}}
\raisebox{0pc}{\includegraphics[width=0.40\textwidth]{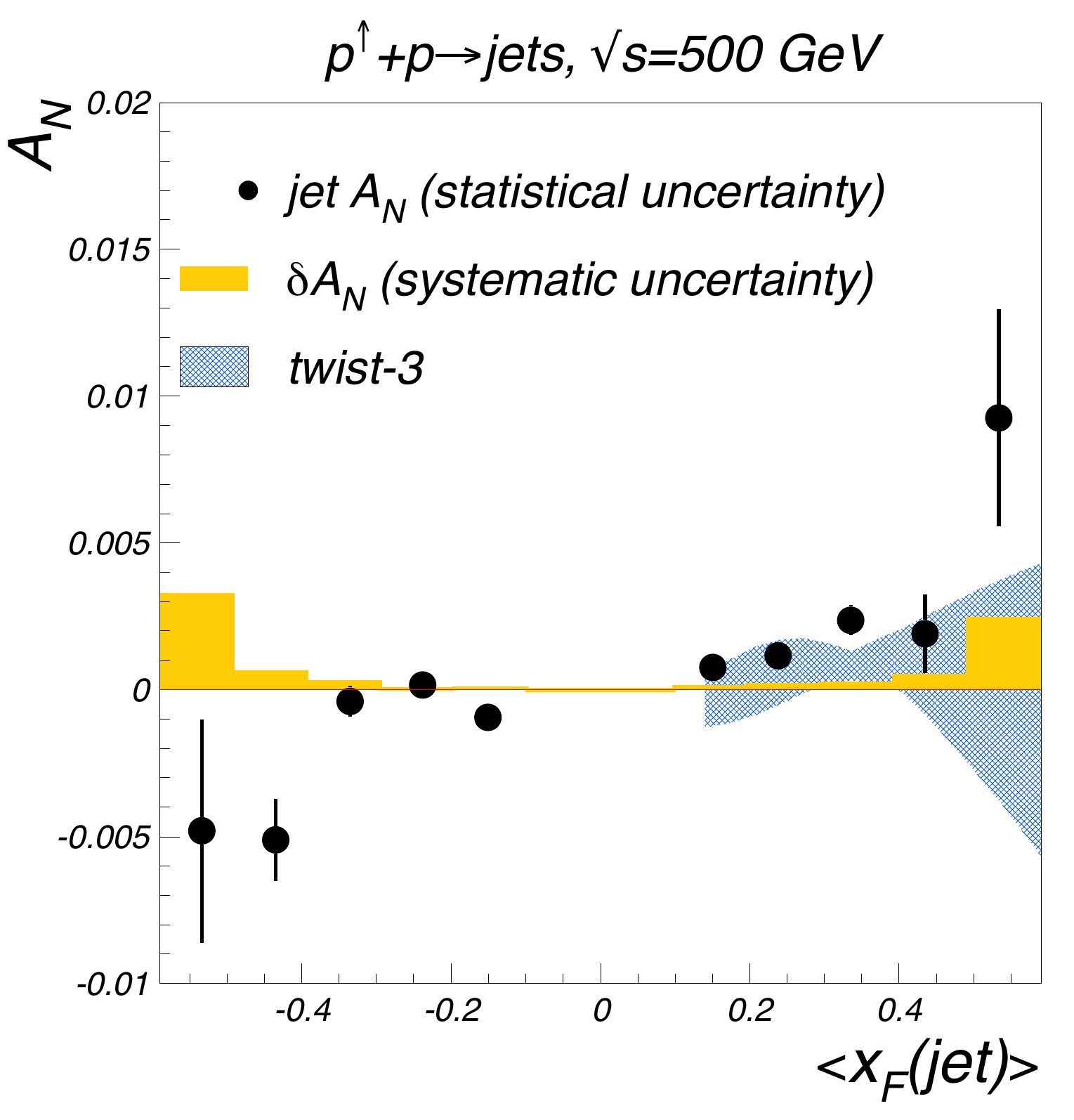}}
\end{center}\vspace*{-1.5pc}
\caption[]{\footnotesize a) $A_N (\pi^0)$ in p$^\uparrow+p$  vs. Feynman $x_F$ for $\sqrt{s}$ indicated~\cite{PXPRD90}. b) $A_N ({\rm jets})$~\cite{ANDYPLB750}.     
\label{fig:ANpi}}
\end{figure}

In practice at RHIC~\cite{PXPRD90} several possible quantities are used to measure $A_N$ (Eq.~\ref{eq:AN})  to cancel luminosity and detector asymmetries, where ${\cal R}$ is the ratio of the luminosities of the two spin states ${\cal L}^\uparrow/{\cal L}^\downarrow$, $N^\uparrow=N_L^\uparrow +N_R^\uparrow$ and $P$ is the beam polarization. 
\begin{equation}
A_N=\frac{1}{P} \frac{d\sigma_L -d\sigma_R}{d\sigma_L +d\sigma_R}= \frac{1}{P} \frac{\sqrt{N_L^\uparrow \cdot N_R^\downarrow}-\sqrt{N_L^\downarrow \cdot N_R^\uparrow}}{\sqrt{N_L^\uparrow \cdot N_R^\downarrow}+\sqrt{N_L^\downarrow \cdot N_R^\uparrow}}= \frac{1}{P} \frac{N^\uparrow -{\cal R} \cdot N^\downarrow}{N^\uparrow +{\cal R} \cdot N^\downarrow}
\label{eq:AN}
\end{equation}  
Transverse single spin asymmetries of forward $\pi^{\pm}$ and $\pi^0$ in p$^\uparrow +$p collisions have been observed from the Argonne ZGS $\sqrt{s}=4.9$ GeV~\cite{KlemPRL36} to RHIC (Fig.~\ref{fig:ANpi}a)~\cite{PXPRD90}. There is very little if any difference as a function of $\sqrt{s}$ and in my personal opinion no clear theoretical understanding of the effect. This leaves room for the experimentalists. 
ANDY, this past year~\cite{ANDYPLB750}, has measured  $A_N$ for forward jets (Fig.~\ref{fig:ANpi}b) and found a much smaller effect. 

A better example is the measurement by PHENIX of $A_N$ of forward ($x_F>0.5$) neutrons in the p$+$A run of 2015, where we decided to request transverse polarization of the protons to see what would happen, although it was primarily a run to measure the high $p_T$ p$+$A baseline using p$+$Al and p$+$Au collisions to vary the nuclear thickness rather than centrality which seems to have problems~\cite{PXdAuPRL116}. The nice feature of the RHIC spin machine is that the bunch to bunch polarization is arranged so that the overall spin effect cancels if an experiment ignores polarization. 

\begin{figure}[!h]
\begin{center}     
\raisebox{1.2pc}{\includegraphics[width=0.43\textwidth]{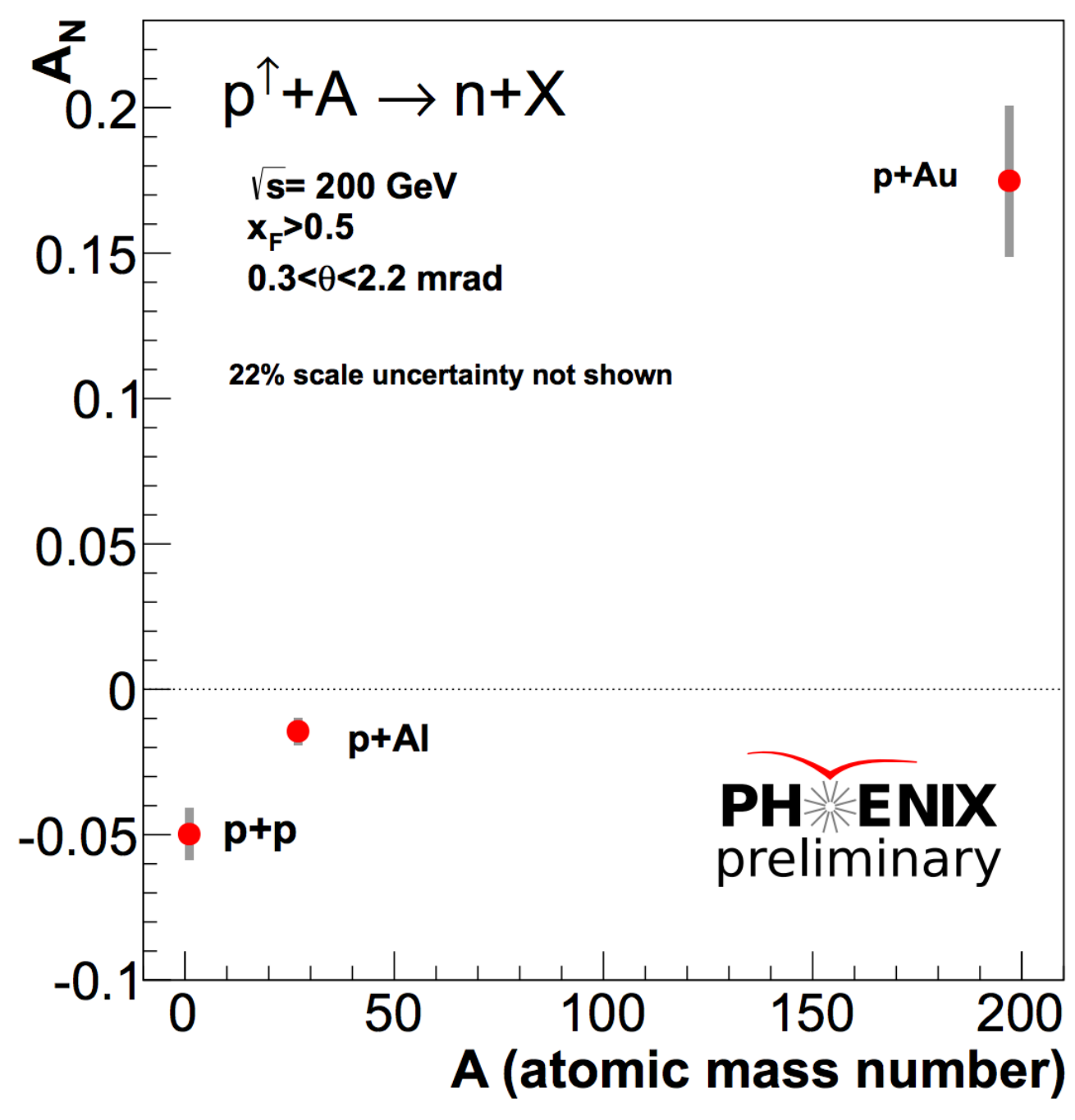}}
\raisebox{0pc}{\includegraphics[width=0.48\textwidth]{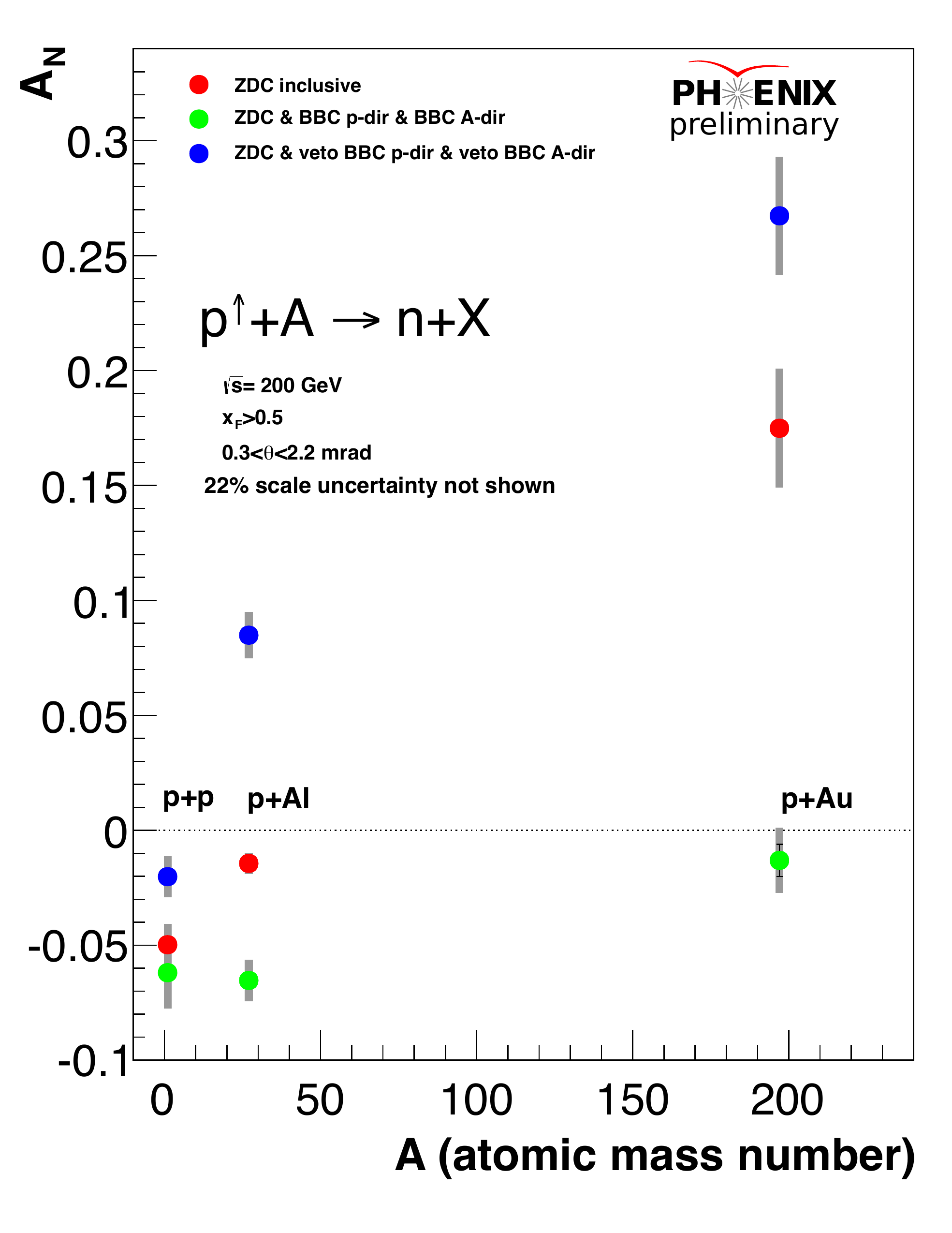}}
\end{center}\vspace*{-1.5pc}
\caption[]{\footnotesize a) $A_N$ vs A for p$^\uparrow +$A$\rightarrow $n+X for A=p, Al and Au~\cite{ItaruMENU}. b) $A_N$ for different BBC activity~\cite{ItaruMENU}.     
\label{fig:ANnA}}\vspace*{-2pc}
\end{figure}

Sure enough, something entirely unexpected happened: a huge A dependence was observed in the single spin asymmetry of forward neutrons (Fig.~\ref{fig:ANnA}a)~\cite{ItaruMENU}.  
Additionally, for all three   
targets (Fig.~\ref{fig:ANnA}b)~\cite{ItaruMENU}) $A_N$ became more positive in the case of Ultra-peripheral or diffractive events for which zero activity was required in both Beam-Beam Counters (BBC) and less positive (dramatically so for Au) when activity in both BBC was required. Now, it's time for the theorists. 
\section{RHIC operation in 2016 and future plans.} 
\begin{figure}[!h]
\begin{center}
\includegraphics[width=0.44\textwidth]{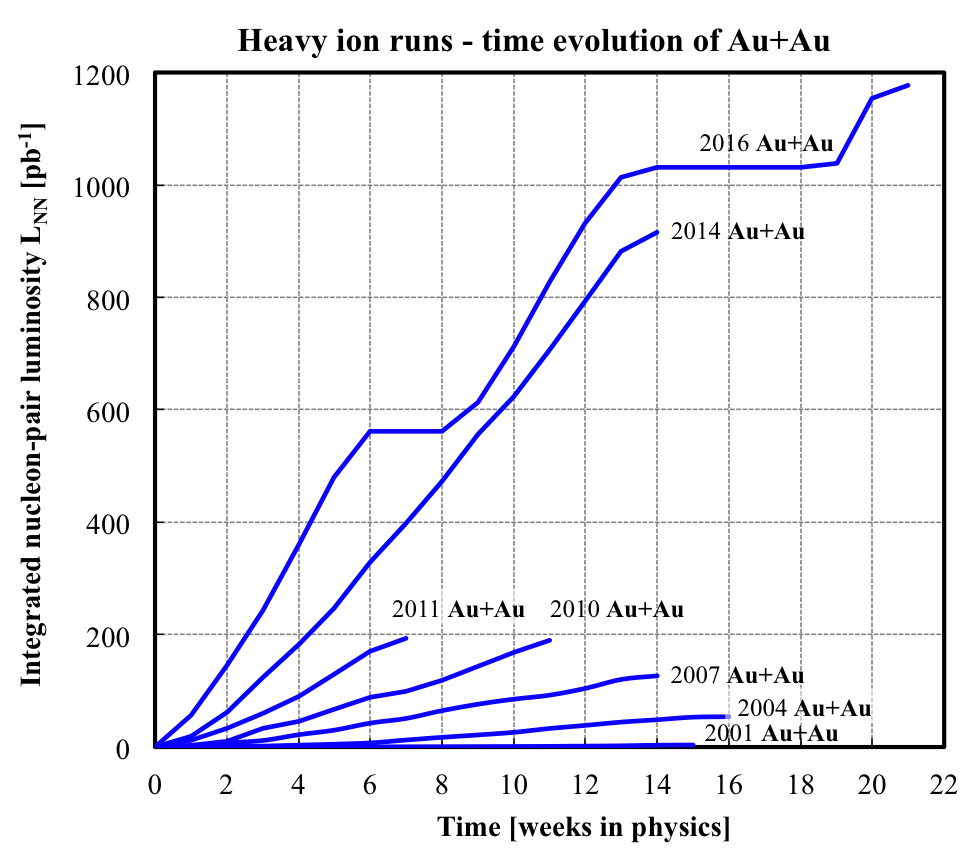}\hspace*{1pc}
\includegraphics[width=0.44\textwidth]{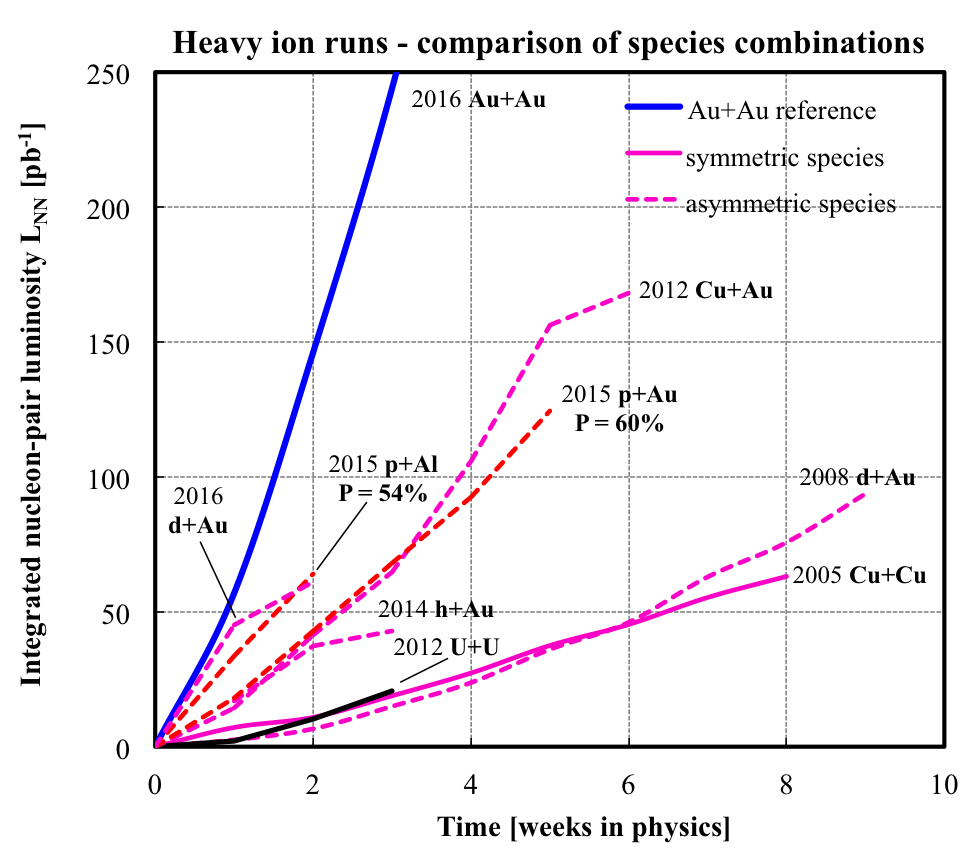}
\end{center}\vspace*{-1.5pc}
\caption[]{\footnotesize A$+$B performance, where the nucleon-pair luminosity is defined as $L_{\rm NN}=A\times B\times L$, where $L$ is the luminosity and $A$, $B$ are the number of nucleons in the colliding species. a)(left) Au$+$Au runs. b) Other species including 2016 d$+$Au and p$^\uparrow +$A runs. Courtesy Wolfram Fischer.}
\label{fig:RHICperf}
\end{figure}

\begin{figure}[!t]
\begin{center}
\raisebox{0.35pc}{\includegraphics[width=0.26\textwidth]{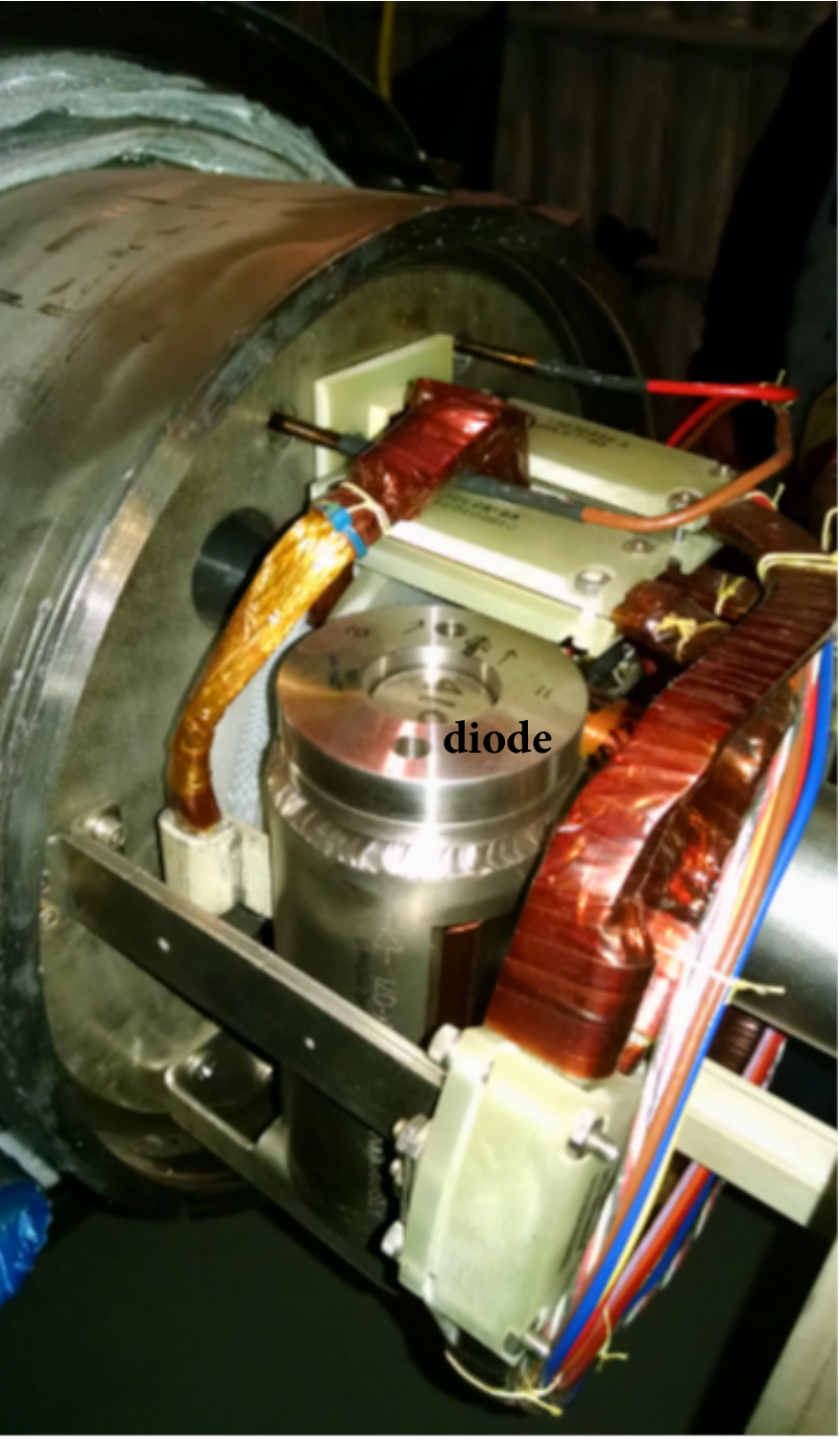}}
\raisebox{0pc}{\includegraphics[width=0.72\textwidth]{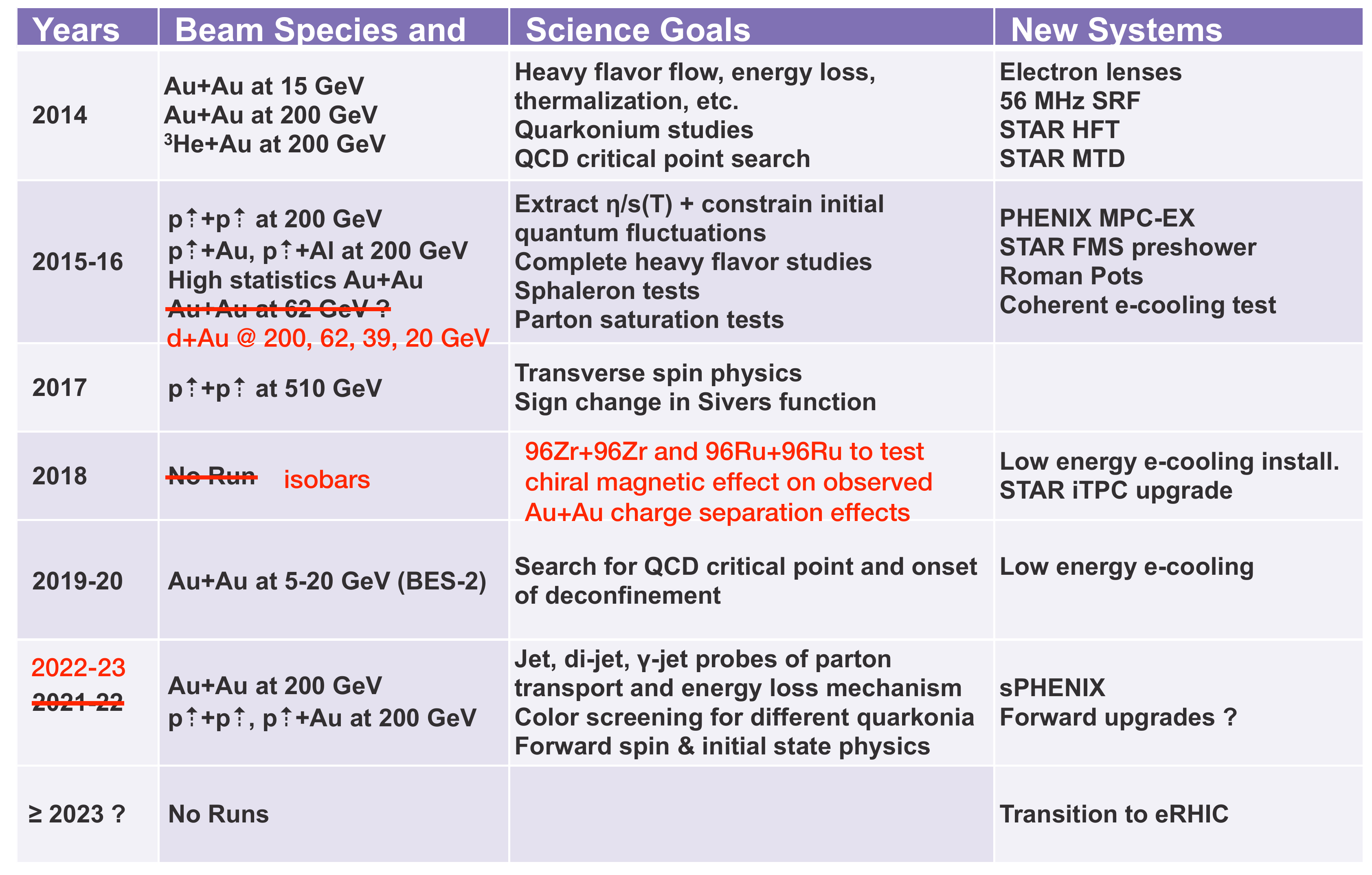}}
\end{center}\vspace*{-1.5pc}
\caption[]{\footnotesize a)(left) Dipole with new quench protection diode before welding. b) (right) New RHIC run schedule with previous entries crossed out. sPHENIX will be installed in 2021.}
\label{fig:RHICrunssched}\vspace*{-1.0pc}
\end{figure}

Physics data taking started on February 8, 2016 (following a blizzard) with a high luminosity 200 GeV Au$+$Au run planned for 10 weeks, primarily for STAR to measure suppression (or enhancement) and flow of $\Lambda_c$, $D$ mesons and Upsilons, to be followed by a d$+$Au beam energy scan for 5 weeks for PHENIX to study the \sqsn dependence of collecivity/flow in small systems for \sqsn=200, 62.4,39 and 19.6 GeV.  The run went very well until a quench protection diode inside a ring dipole magnet malfunctioned on March 18 and had to be replaced. This involved warming up of a sector of RHIC magnets, cutting open the dipole with the faulty diode, removing and replacing it, closing up the dipole, and cooling down the ring. The run resumed on April 6, was able to be extended a few weeks more than planned thanks to lower electrical costs, and ended on June 27 with both PHENIX and STAR largely meeting their luminosity goals. Apart from the malfunction, the RHIC machine operated better than ever, with higher luminosity and a flatter luminosity profile. The luminosity performance is shown in Fig.~\ref{fig:RHICperf}. Note that the 2016
d$+$Au shows only 2 weeks, which were at \sqsn=200 and 62.4 GeV. The \sqsn= 39 and 19.4 GeV runs for the next two weeks only added 7\% additional events. The replaced quench protection diode is shown in Fig.
~\ref{fig:RHICrunssched}a. 

This year's run ended data taking by PHENIX which spanned 16 years. This year was also the 25th anniversary of the PHENIX collaboration which started in August 1991 when 3 proposals to the RHIC Program Advisory Committee were merged by the Associate Laboratory Director, Mel Schwartz, into an experiment ``to study electrons and photons emerging from the Quark Gluon Plasma''.  

The proposed schedule for runs and new equipment from 2014 to $\geq 2023$? shown in Fig.~\ref{fig:RHICrunssched}b has changed since last year thanks to news from the DOE that operations money in years when the machine doesn't operate may be reduced, unlike the CERN schedule. The principal change is that there will be a 2018 run with collisions of isobars, $^{96}_{40}$Zr + $^{96}_{40}$Zr compared to $^{96}_{44}$Ru + $^{96}_{40}$Ru, to understand whether the charge separation of anisotropic flow $v_2$ of $\pi^+$ and $\pi^-$ observed by STAR in Au$+$Au~\cite{STARPRL114}, the so-called Chiral Magnetic Effect, will be different for the different $Z$, hence due to the strong electromagnetic field in the nuclear collisions, or will remain unchanged for collisions of nuclei with the same number of nucleons. The year 2021 is reserved for the installation of sPHENIX, with data runs taking place in 2022 and 2023. Beyond 2023 is anybody's guess. However, Tim Hallman, Associate Director for Nuclear Physics in the DOE Office of Science, formerly Group leader of the BNL-STAR group and spokesperson of the STAR experiment, left us with some excellent advice at this year's RHIC User's meeting~\cite{HallmanAUM16}:``An important challenge is charting and being able to follow a course to this future which realizes expected scientific return on existing investment and does not leave important science discoveries `on the table'--forever perhaps.'' 
\subsection{sPHENIX progress}
\begin{figure}[!t]
\begin{center}
\raisebox{0pc}{\includegraphics[width=0.58\textwidth]{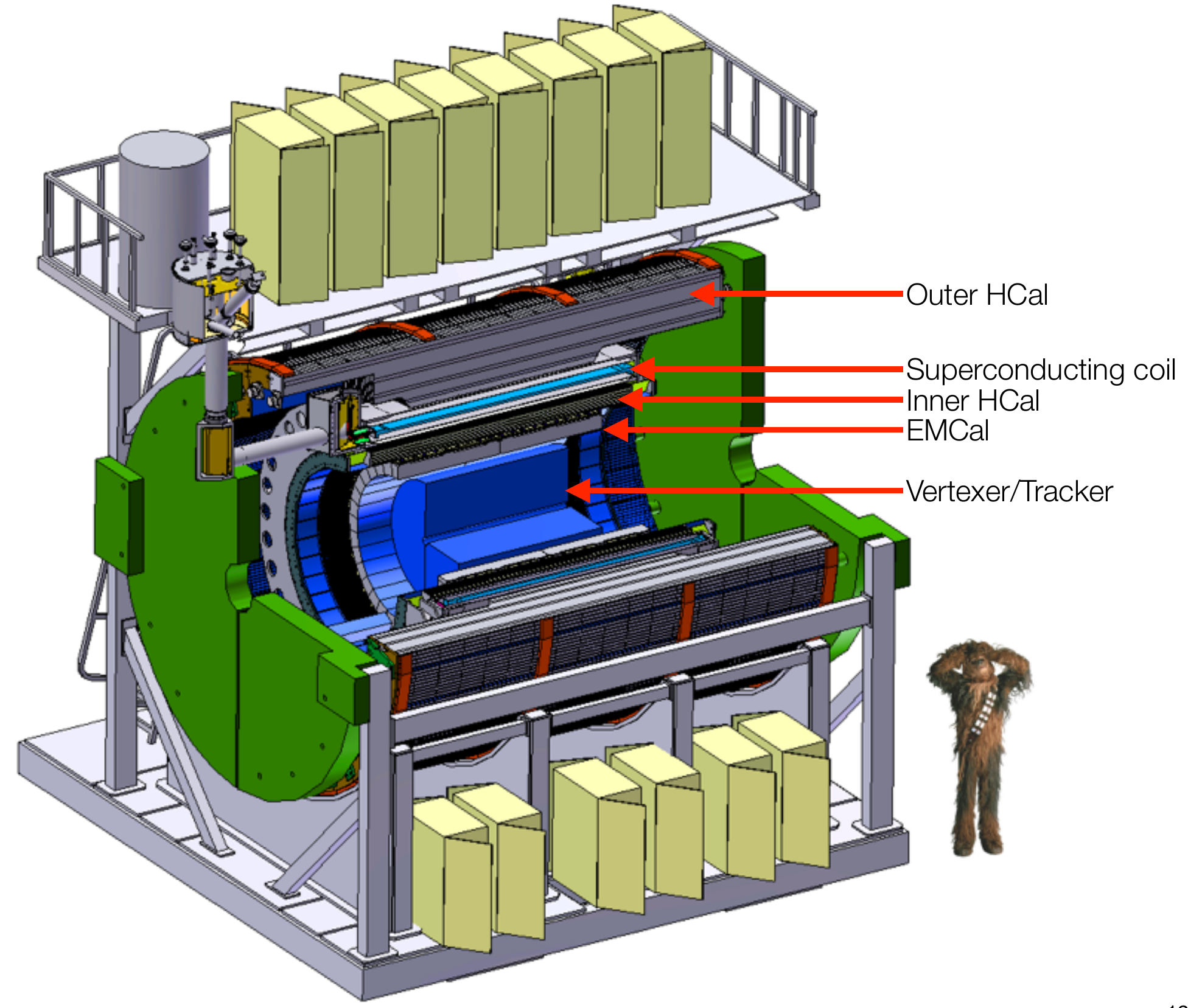}}
\raisebox{0pc}{\includegraphics[width=0.38\textwidth]{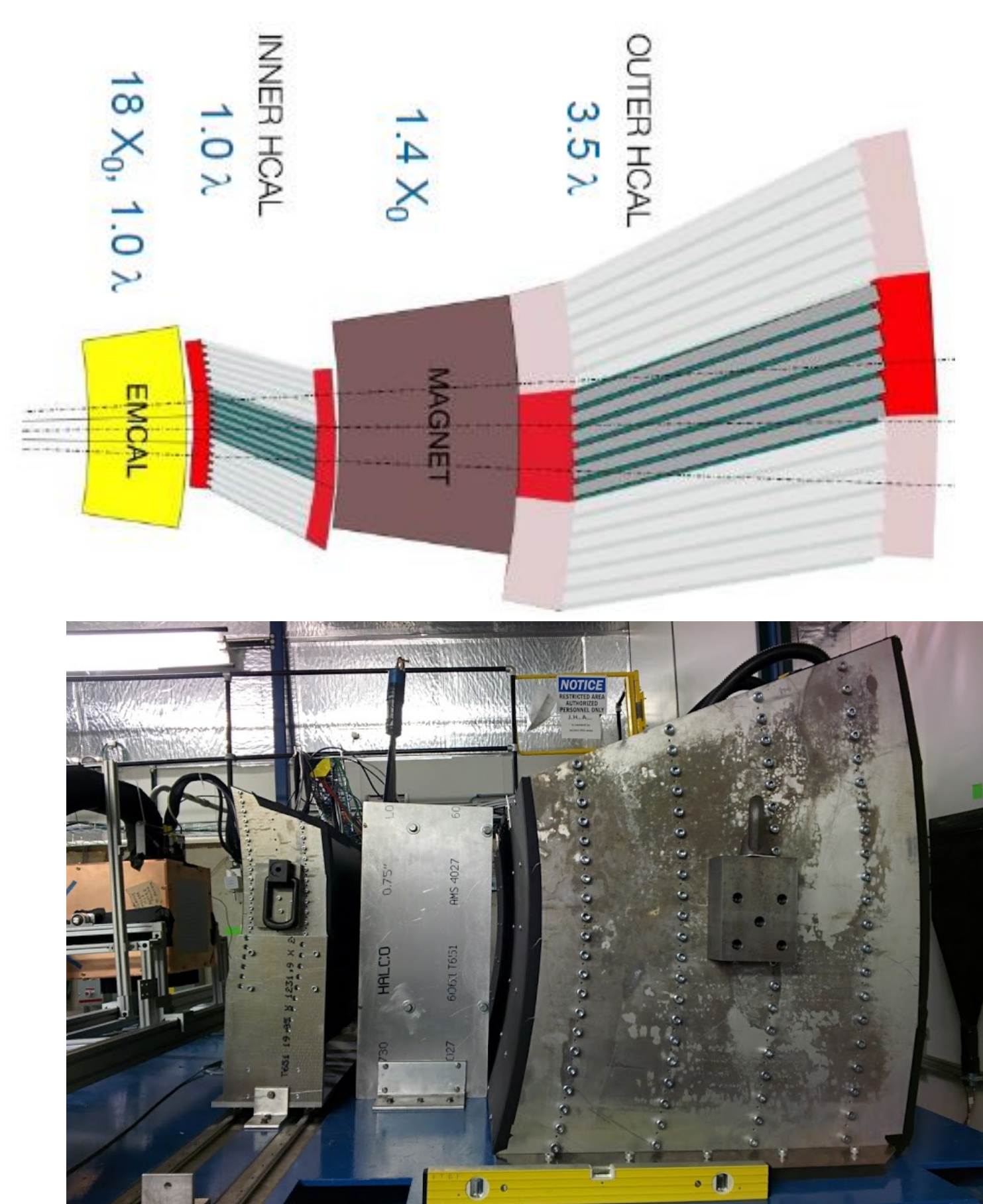}}
\end{center}\vspace*{-1.5pc}
\caption[]{\footnotesize a)(left) sPHENIX engineering design. b) (right) Azimuthal wedge of EMCal, HCal  inside solenoid, solenoid coil, Outer HCal: (top) design; (bottom) beam test.}
\label{fig:sPHENIX}\vspace*{-1.0pc}
\end{figure}

This past year, the sPHENIX project became a formal collaboration, with an inagural meeting at Rutgers University in December 2015 at which Bylaws were approved, and regular meetings since then. Dave Morrison of BNL and Gunther Roland of MIT were elected as co-spokespersons. The objective of the experiment is to make precision measurements of Upsilon suppression and quenching of jets and $b$-quark jets up to $\pt \approx 50$ GeV/c in a hadron calorimeter to probe the structure and properties of the ``perfect liquid'' \QGP. Progress so far this year has been the first cool-down and excitation of the sPHENIX superconducting solenoid at low current (100 Amperes) which generated the expected 256 Gauss magnetic field---the Babar/Ansaldo magnet  works like new. Improved engineering design of the whole detector (Fig.~\ref{fig:sPHENIX}a), construction and testing of a wedge of the proposed calorimeter in a test beam at FERMILAB (Fig.~\ref{fig:sPHENIX}b), as well as  design of the charged particle tracking are ongoing.
\vspace*{-1.5pc}
\section{More Physics by Press Release}
In previous ISSP meetings and proceedings, I have objected to physics by press release. This year
BNL made two such releases: one that's deserved, in my opinion, and one that isn't. \vspace*{-1.0pc}
\subsection{The one that's deserved}\vspace*{-1.0pc}
\begin{figure}[!h]
\begin{center}
\raisebox{0pc}{\includegraphics[width=0.58\textwidth]{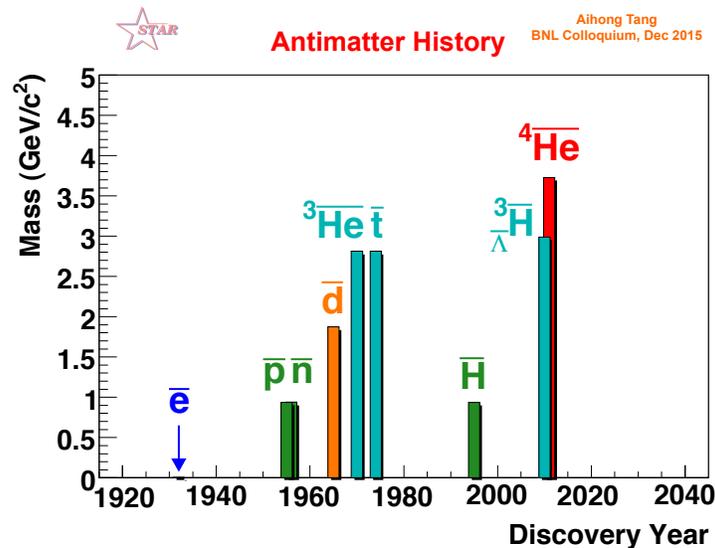}}
\end{center}\vspace*{-2.5pc}
\caption[]{\footnotesize ``Antimatter'' Discovery History. Courtesy Aihong Tang (BNL).}
\label{fig:antimatter}\vspace*{-1.0pc}
\end{figure}

On November 4, 2015, BNL posted the press release: \href{https://www.bnl.gov/newsroom/news.php?a=11786}{``Physicists Measure Force that makes Antimatter Stick Together''}. This clever experiment by STAR~\cite{NatureHBTpbar} used the method of Hanbury-Brown Twiss Correlations to measure the force between two anti-protons emitted in \sqsn=200 GeV Au$+$Au collisions at RHIC. Antimatter production has a long history following Dirac~\cite{DiracPositron} (Fig.~\ref{fig:antimatter}) 
but only since 2010 has the field shifted to production in A$+$A collisions at RHIC~\cite{STAR2010} and LHC~\cite{ALICENatureP2015}, where typically for $\sqsn\gsim 200$ GeV as many anti-baryon  as baryons are created in large numbers, $\mu_B\rightarrow 0$ .  

I was a bit nervous showing Fig.~\ref{fig:antimatter} in my lecture because I expected Prof. Zichichi, who discovered the antideuteron ($\bar{d}$)~\cite{MassamNCA63}\footnote{The $\bar{d}$ was also observed at BNL~\cite{Dorfandbar} in an experiment looking for fractionally charged constituent quarks. Ref.~\cite{MassamNCA63} was submitted for publication two months before Ref.~\cite{Dorfandbar} but appeared in print three months after.}, to comment (he did) and to explain that antiparticles do not constitute antimatter: there must be some sort of nuclear ``antiglue'' to bind antiprotons to antineutrons. ``If the antideuteron did not exist, nothing but light antihydrogen could exist: farewell anti-water and farewell all forms of antimatter''~\cite{ZichichiCC2009}. 

\begin{figure}[!h]
\begin{center}
\raisebox{0.4pc}{\includegraphics[width=0.48\textwidth]{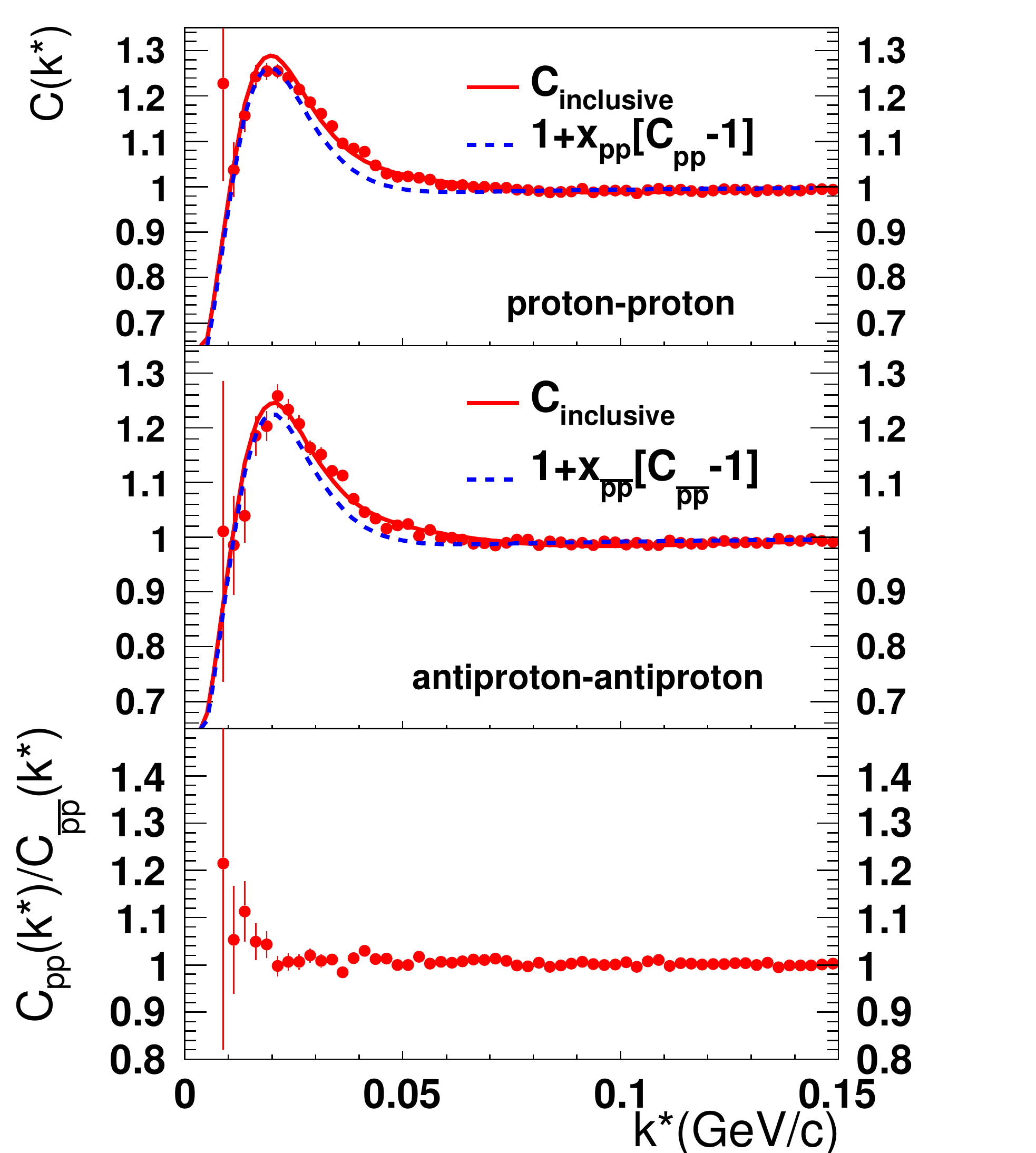}}
\raisebox{0pc}{\includegraphics[width=0.42\textwidth]{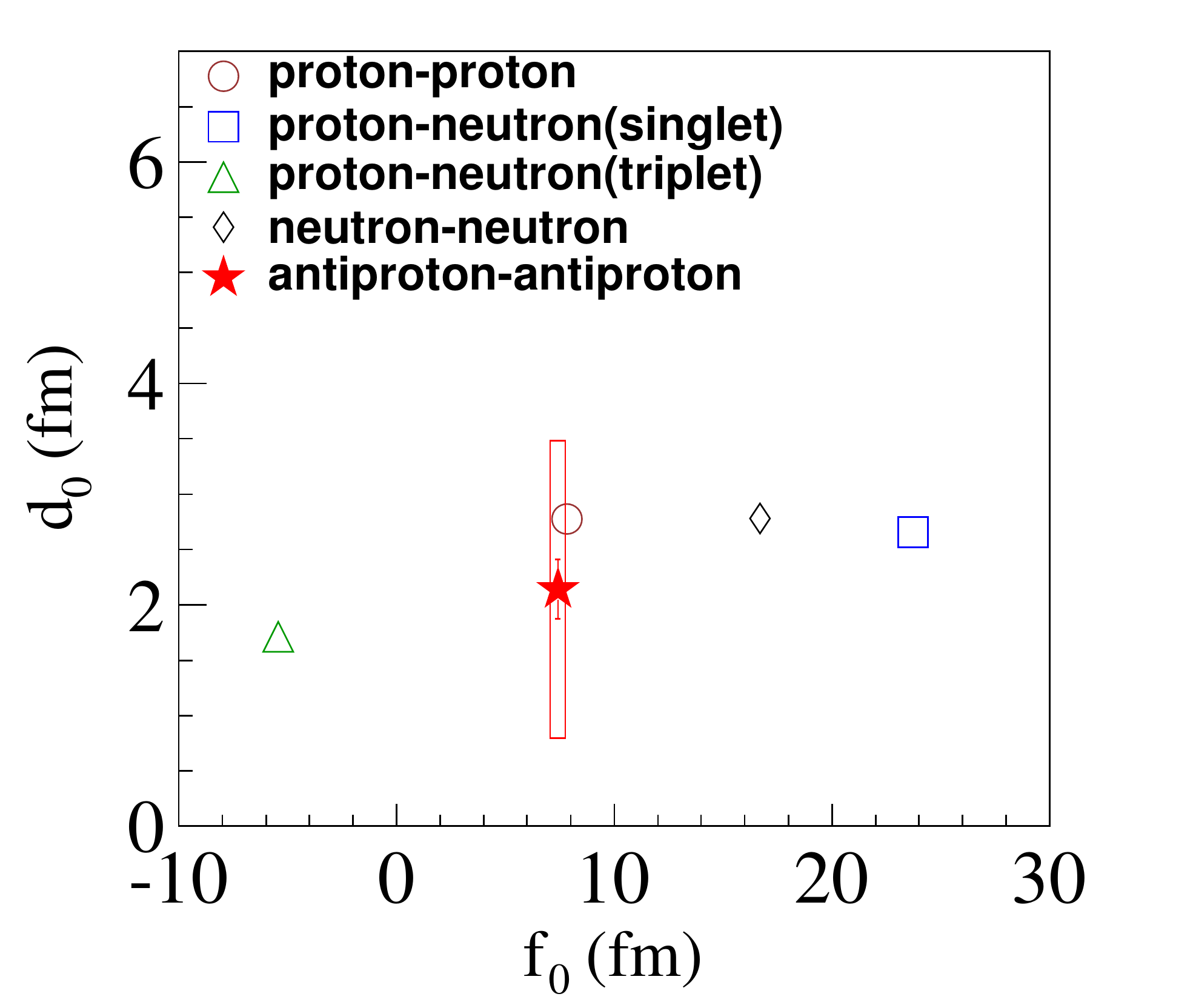}}
\end{center}\vspace*{-1.5pc}
\caption[]{\footnotesize a)(left) Correlation functions of $p+p$, $\bar{p}+\bar{p}$ and their ratio.  b) (right) Calculated d$_0$ and f$_0$ for $\bar{p}+\bar{p}$ compared to $p+p$, $p+n$ and $n+n$ measurements~\cite{NatureHBTpbar}. }
\label{fig:Cpp}\vspace*{-1.0pc}
\end{figure}

In Ref.~\cite{NatureHBTpbar} the force between two antiprotons was represented in terms of the low energy scattering length (f$_0$) and the effective range (d$_0$), which are related to the $s$-wave scattering phase shift $\delta_0$, and the momentum difference $k$ of the two $\bar{p}$'s. These can be derived from 
the $\bar{p}+\bar{p}$ correlation function and compared to the $p+p$ correlation. Figure~\ref{fig:Cpp}a shows that the correlation functions for  $p+p$ and $\bar{p}+\bar{p}$ appear to be identical as shown by their ratio $\approx 1.0$. In Fig.~\ref{fig:Cpp}b the scattering length and effective range for  $\bar{p}+\bar{p}$ calculated from the correlation function are in excellent agreement with the matter  measurements. \vspace*{-0.5pc}
\subsection{The one that's not deserved}
On December 7, 2015, BNL posted the press release: \href{https://www.bnl.gov/rhic/news2/news.asp?a=1797&t=pr}{``RHIC Particle Smashups Find that Shape Matters''}, with the sub-heading,``Scientists colliding football- and sphere-shaped ions discover evidence supporting a paradigm shift in the birth of the quark-gluon plasma.'' It goes on, ``scientists have come to a new understanding of how particles are produced in these collisions. This understanding represents a paradigm shift consistent with the presence of a saturated state of gluons, super-dense fields of the glue-like particles that bind the building blocks of ordinary matter.'' Of course what they left out in the press release is that in the actual publication~\cite{STARUUPRL} the Constituent Quark Model works as well, in fact better, than ``the saturated state of gluons'' (CGC-IPGlasma model), and the paradigm shift---that the number of collisions \Ncoll was not relevant for soft particle production in multiplicity distributions  but the \Nqp (number of constituent quark participants) worked---had been published in 2014~\cite{PXPRC89}. In fact, I addressed the issue---that the predicted sharp reduction of $v_2$ in central U$+$U collisions because of the predicted dominance of the tip-to-tip configuration from  \Ncoll dominance was WRONG---at ISSP2014 and in the proceedings, which section I repeat here.  
\subsubsection{$\mathbf{v_2}$ in U$+$U collisions and constituent-quark participants (ISSP2014)} 
\begin{figure}[!thb]
   \begin{center}
a)\raisebox{0.4pc}{\includegraphics[width=0.38\linewidth]{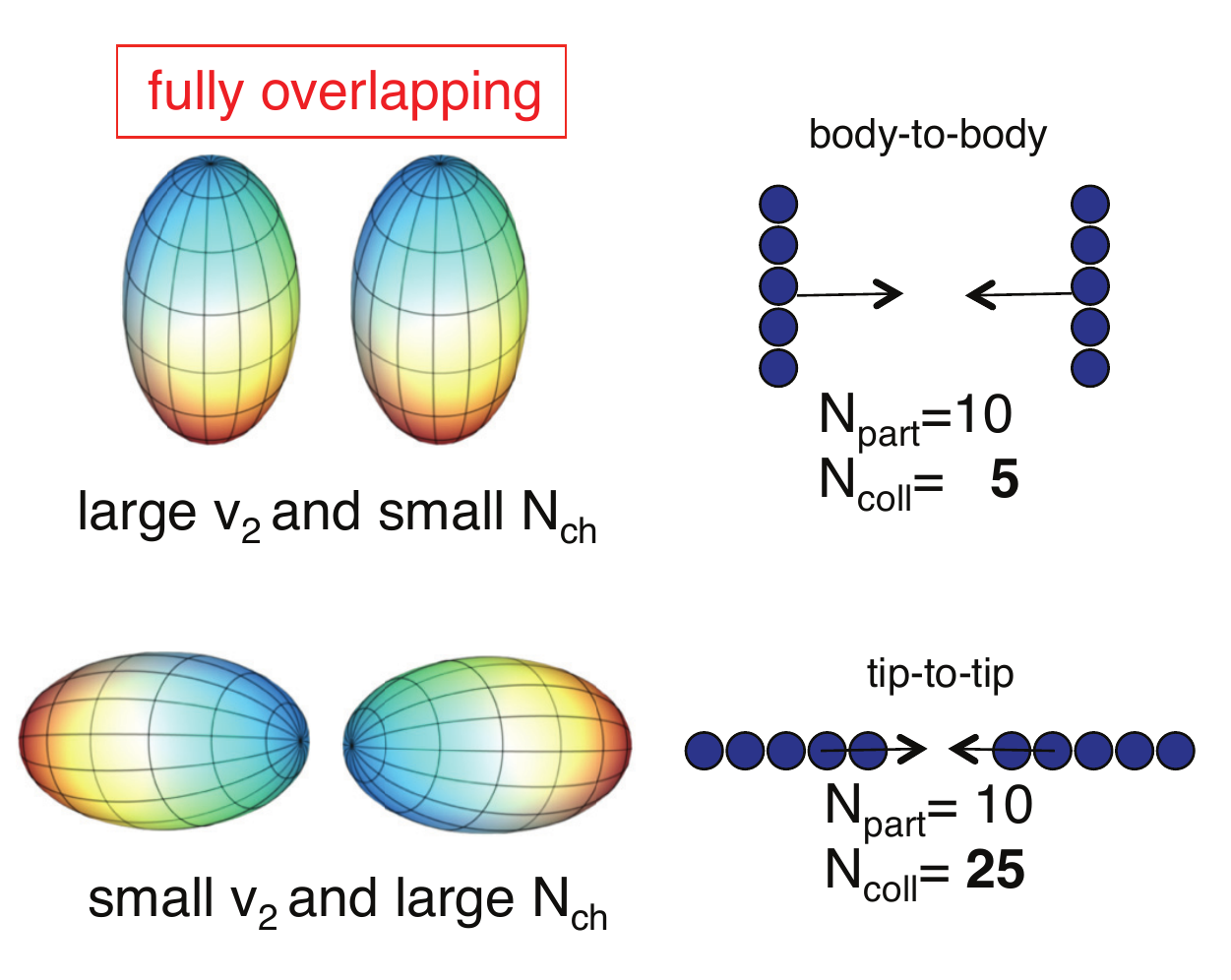}}\hspace*{1.0pc}
b)\raisebox{0.0pc}{\includegraphics[width=0.54\linewidth]{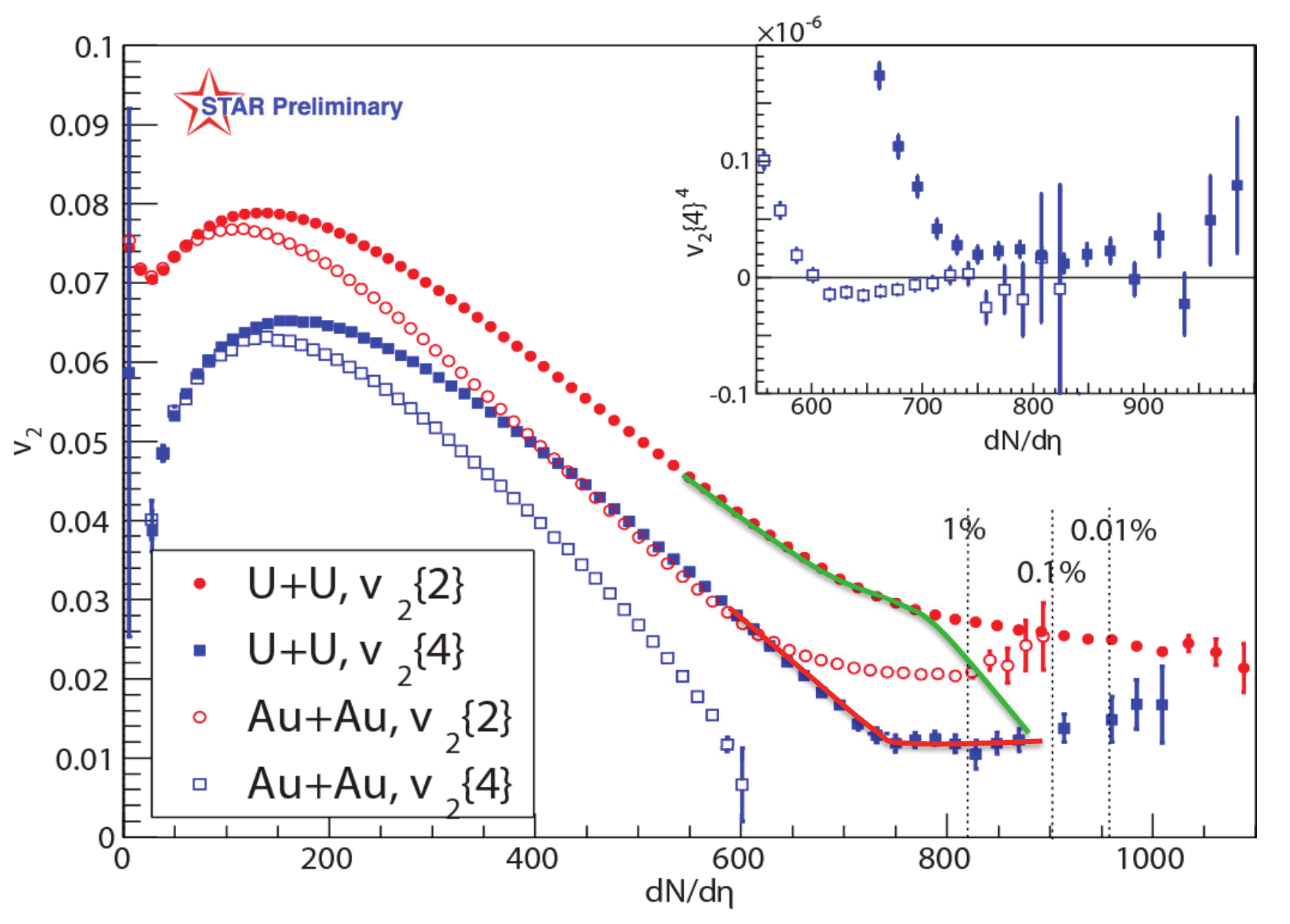}}
\end{center}\vspace*{-1.0pc}
\caption[]{\footnotesize (a) Body-to-body and tip-to-tip configurations in U+U collisions with zero impact parameter. The different relation of \Npart to \Ncoll is sketched next to each configuration.
(b) STAR measurements of $v_2$ in Au$+$Au and U$+$U at $\sqsn\approx$ 200 GeV as a function of $d\Nch/d\eta$ with upper percentiles of centrality for U$+$U indicated by vertical dashed lines~\cite{STARUUPRL}.
\label{fig:UUv2}}
\end{figure}

Because Uranium nuclei are prolate spheroids, there is the interesting possibility of large $v_2$ in body-to-body central collisions which have a significant eccentricity and almond shape (Fig.~\ref{fig:UUv2}a).  
Based on the assumption that the \Ncoll is relevant to describe the $d\Nch/d\eta$ distribution in U$+$U collisions, it was predicted that for the highest $d\Nch/d\eta$ (the most central collisions) the tip-to-tip configuration with much larger \Ncoll and small eccentricity (small $v_2$) would overtake the body-to-body configuration with large eccentricity corresponding to large $v_2$.  

This led to two predictions: i) the tip-to-tip configuration would be selected by the most central collisions~\cite{KuhlmanHeinzPRC72}; ii) these most central collisons would see a sharp decrease in $v_2$ with increasing $d\Nch/d\eta$~\cite{Filip--NuXuPRC80,VoloshinPRL105} called a cusp. This sharp decrease---represented by the bent line on the topmost U$+$U data (filled circles) in Fig.~\ref{fig:UUv2}b (not shown in Ref.~\cite{STARUUPRL})---is not observed. As discussed previously, this is because the \Ncoll term is not relevant for $d\Nch/d\eta$ distributions, which also argues against the method proposed in Ref.~\cite{KuhlmanHeinzPRC72} to select the tip-to-tip configuration. 

\section{New results on Constituent Quark Participants}  
Before presenting the new results, I briefly review the PHENIX2014~\cite{PXPRC89} number of constituent quark participants (NQP) model of mid-rapidity \Et and $d\Nch/d\eta$ distributions. 
\subsection{The PHENIX2014~\cite{PXPRC89} NQP model}
	The massive constituent-quarks~\cite{MGMPL8,MuellerNPA527,MorpurgoRNC33}, which form mesons and nucleons (e.g. a proton=$uud$), are relevant for static properties and soft physics with $p_T\lsim1.4$ GeV/c. They are complex objects or quasiparticles~\cite{ShuryakNPB203} made of the massless partons (valence quarks, gluons and sea quarks) of DIS~\cite{DIS2} such that the valence quarks acquire masses $\approx 1/3$ the nucleon mass with radii $\approx 0.3$ fm when bound in the nucleon. With  finer  resolution one can see inside the bag to resolve the massless partons which can scatter at large angles according to QCD. At RHIC, hard-scattering starts to be visible as a power law above soft (exponential) particle production only for $p_T>$ 1.4 GeV/c at mid-rapidity~\cite{PXpi0PRD}, where $Q^2=2p_T^2=4$ (GeV/c)$^2$ which corresponds to a distance scale (resolution) $<0.1$ fm.
	
\begin{figure}[hbt] 
      \centering
 \includegraphics[width=0.48\linewidth]{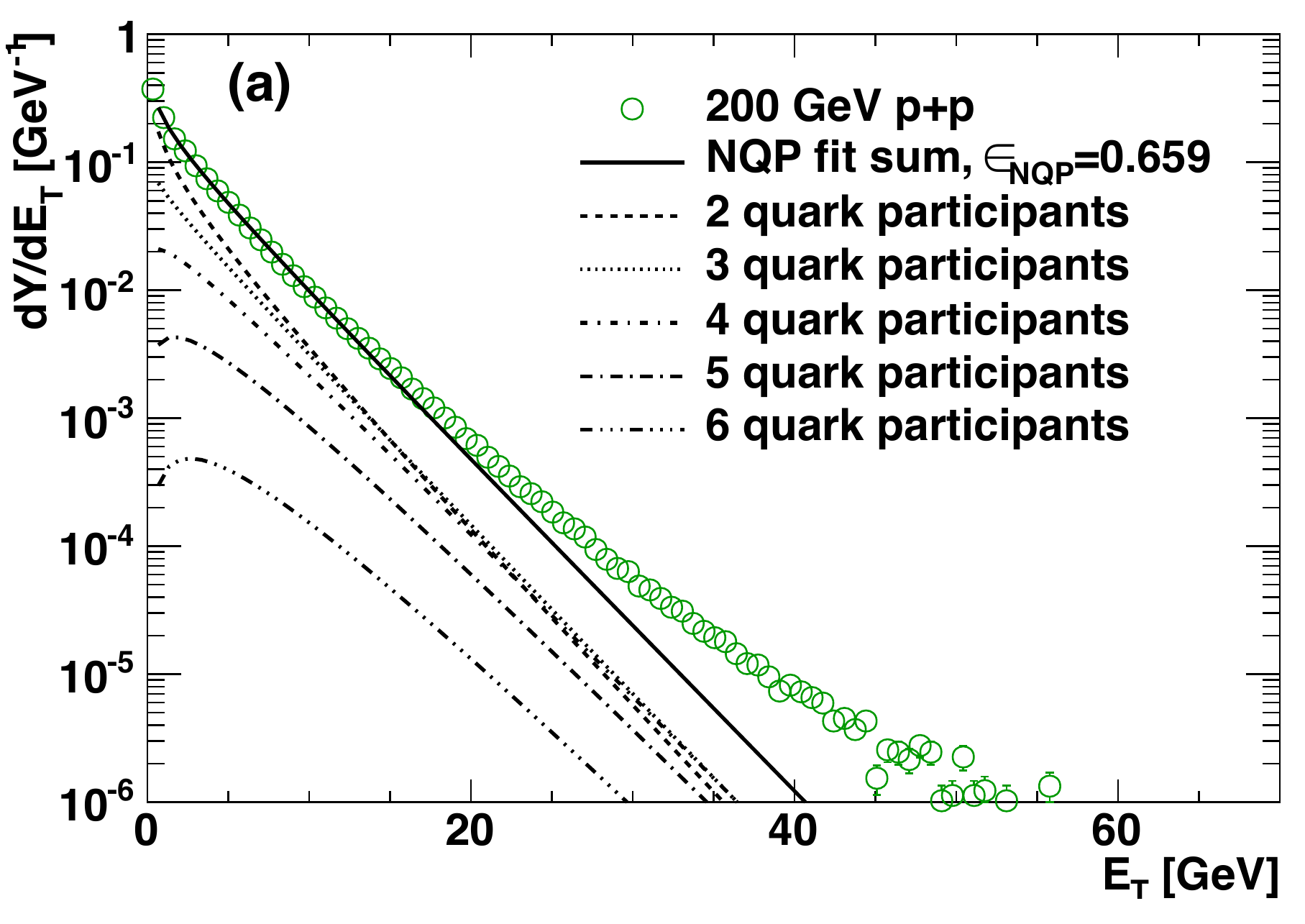}
      \includegraphics[width=0.48\linewidth]{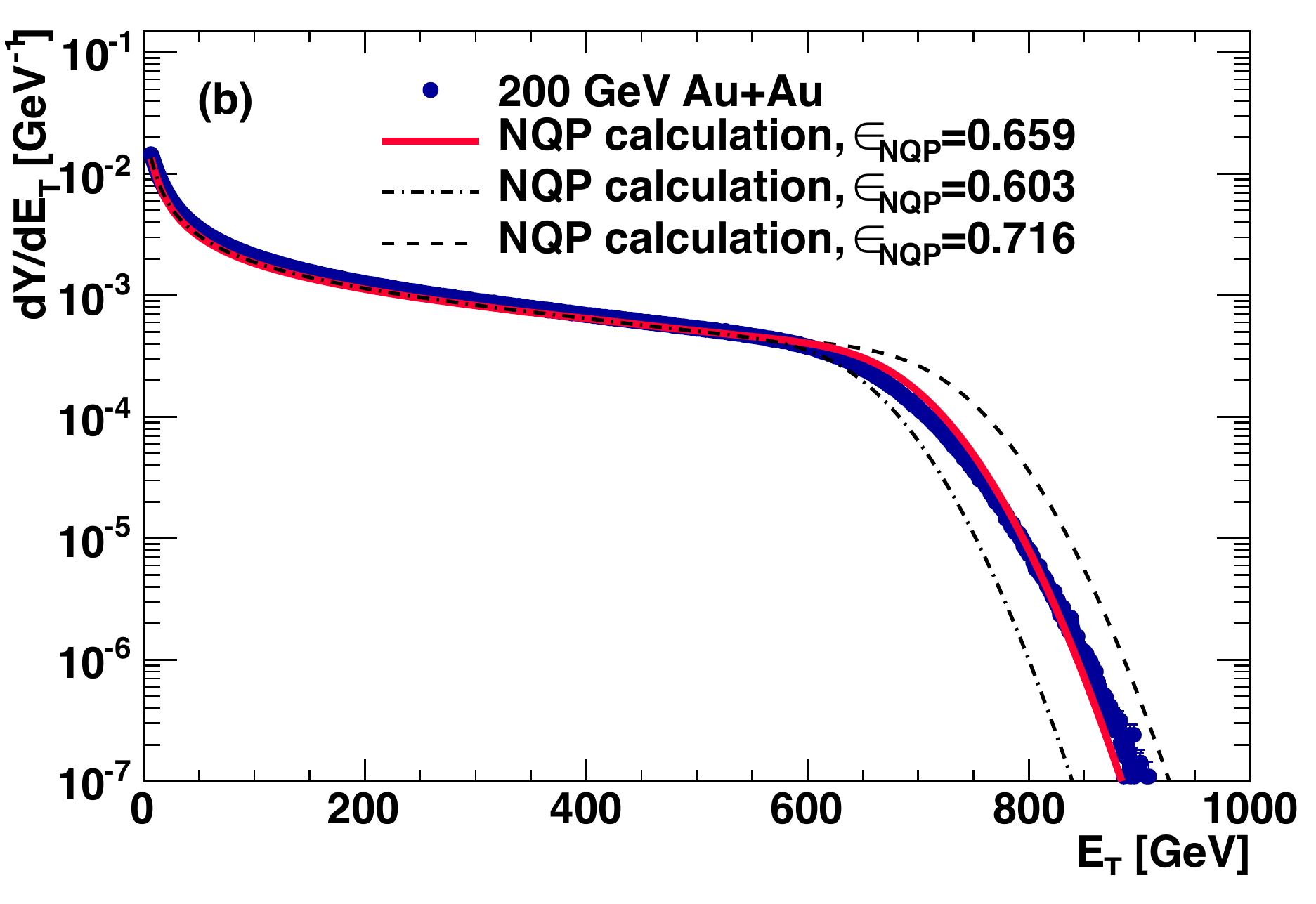}\vspace*{-0.5pc} 
      \caption[]{\footnotesize PHENIX2014~\cite{PXPRC89} method for $\Et\equiv d\Et/d\eta|_{y=0}$ distributions at $\sqsn=200$ GeV:
      a) Deconvolution fit to the $p$+$p$ \Et distribution for $\Et<13.3$ GeV for $\epsilon_{\rm NQP}=1-p_{0_{\rm NQP}}=0.659$ calculated in the Number of Quark Participants or \Nqp model. Lines represent the properly weighted individual \Et distributions for the underlying 2,3,4,5,6 constituent-quark participants plus the sum.  b) Au+Au \Et distribution compared to the NQP calculations using the central $1-p_0=0.647$ and $\pm 1\sigma$ variations of $1-p_0=0.582,0.712$ for the probability $p_0$ of getting zero \Et on a $p$+$p$ collision with resulting Quark-Participant efficiencies $\epsilon_{\rm NQP}=0.659,0.603,0.716$, respectively.        \label{fig:PXppg100}} \vspace*{-0.5pc}
   \end{figure}

	The PHENIX2014~\cite{PXPRC89} calculation starts by generating the positions of the nucleons in each nucleus of an A$+$B collision by the standard method. Then the spatial positions of the three quarks are  generated around the position of each nucleon using the proton charge distribution corresponding to the Fourier transform of the form factor of the proton~\cite{HofstadterRMP28,HofstadterRMP30}:
\begin{equation}
   \rho^{\rm proton}(r) = \rho^{\rm proton}_{0} \times \exp(-ar),
   \label{eq:Hofstadterdipole}
\end{equation}
where $a = \sqrt{12}/r_{m} = 4.27$ fm$^{-1}$ and 
$r_{m}=0.81$ fm is  
the r.m.s radius of the proton weighted according to charge~\cite{HofstadterRMP28} 
\begin{equation}
r_{m}=\int_0^\infty r^2 \times 4\pi r^2 \rho^{\rm proton}(r) dr \qquad .
\label{eq:rmsintegral}
\end{equation}
The corresponding proton form factor is the Hofstadter dipole fit~\cite{HandMillerWilson} now known as the standard dipole~\cite{BernauerMainzPRC90}:
\begin{equation}
G_E(Q^2)=G_M(Q^2)/\mu=\frac{1}{(1+\frac{Q^2}{0.71 {\rm GeV}^2})^2} \label{eq:HofstadterdipoleFF}
\end{equation}
where $G_E$ and $G_M$ are the electric and magnetic form factors of the proton, $\mu$ is its magnetic moment and $Q^2$ is the four-momentum-transfer-squared of the scattering.
The inelastic $q+q$ cross section $\sigma^{\rm inel}_{q+q}=9.36$mb at \sqsn=200 GeV was derived from the $p+p$ \Nqp Glauber calculation by requiring the  calculated $p+p$ inelastic cross section to reproduce the measured $\sigma^{\rm inel}_{N+N}=42$ mb cross section,  and then used for the Au$+$Au (and $d+$Au-not shown) calculations (Fig.~\ref{fig:PXppg100})~\cite{PXPRC89}. 

\begin{figure}[!hbt] 
      \centering
\raisebox{0.0pc}{\includegraphics[width=0.54\linewidth]{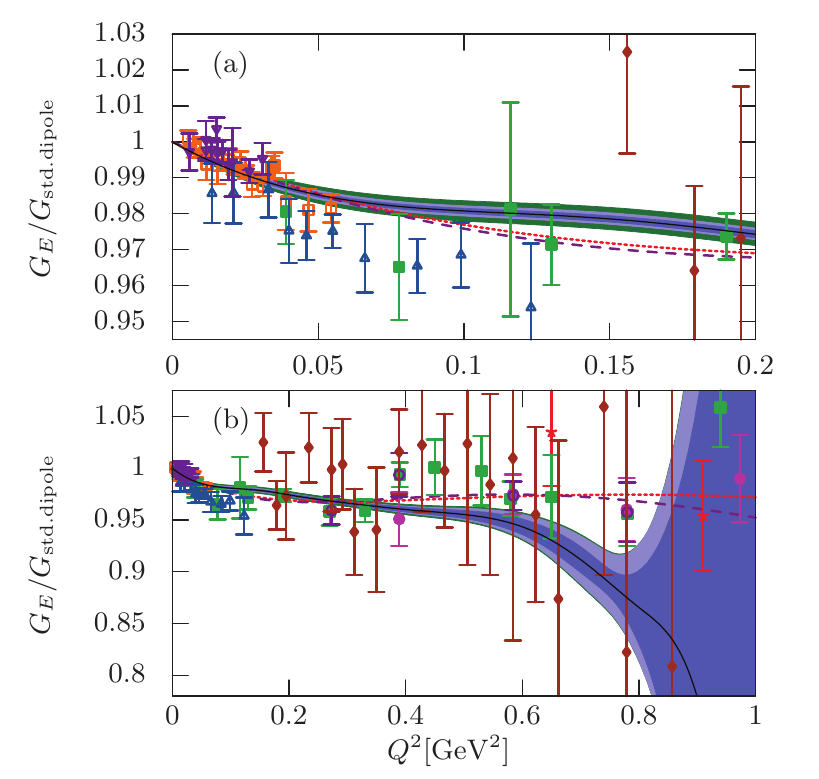}}
\raisebox{0.0pc}{\includegraphics[width=0.42\linewidth]{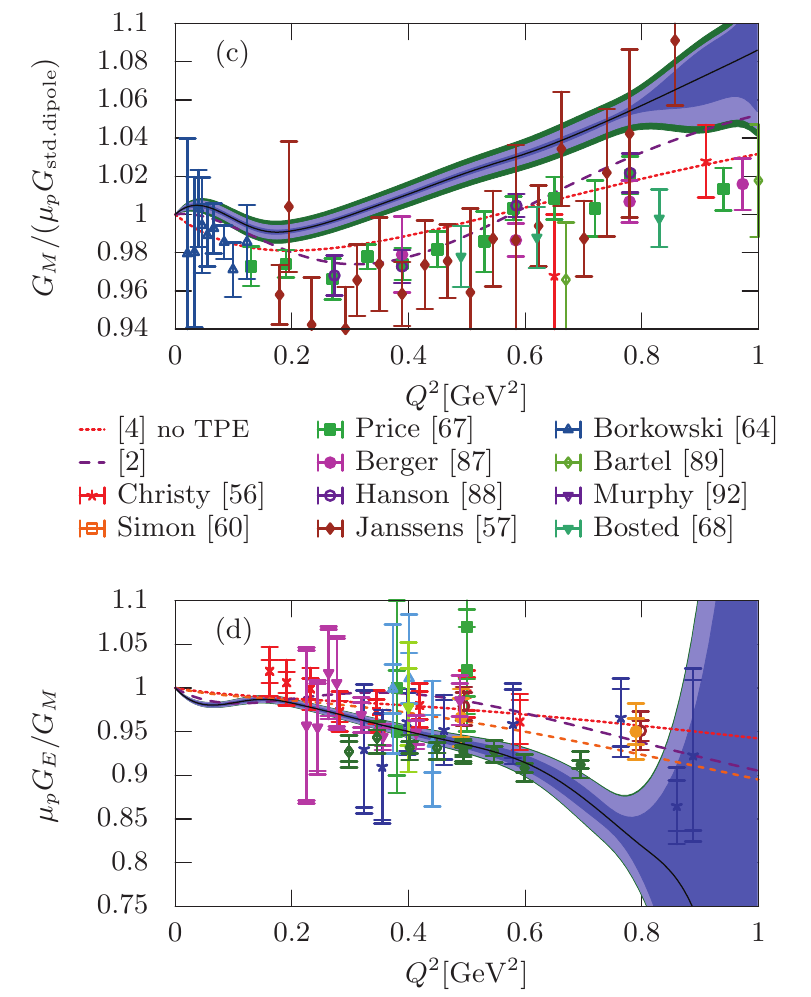}}\vspace*{-0.5pc} 
      \caption[]{\footnotesize The form factors $G_E$ and $G_M$, normalized to the standard dipole, and $G_E/G_M$, compared to fits, with the dark region being the best fit to the new Mainz data~\cite{BernauerMainzPRC90}. }\vspace*{-0.5pc}
      \label{fig:Mainz}
   \end{figure}

People sometimes ask why we use Hofstadter's 60 year old measurements when there are more modern measurements which give a different proton r.m.s charge radius~\cite{BernauerMainzPRC90}	, which is not computed from Eq.~\ref{eq:rmsintegral} but merely from the slope of the form factor at $Q^2=0$. The answer is given in Fig.~\ref{fig:Mainz} which shows how all the measurements of $G_E(Q^2)$ and $G_M(Q^2)$ for $Q^2\leq  1$ GeV$^2$ agree with the ``standard dipole'' (Eq.~\ref{eq:HofstadterdipoleFF}) within a few percent and in all cases except Fig.~\ref{fig:Mainz}d agree better than the Mainz fit. 
\subsection{Improved method of generating constituent quaris}
A few months after PHENIX2014~\cite{PXPRC89} was published, it was pointed out to us that our method did not preserve the radial charge distribution (Eq.~\ref{eq:Hofstadterdipole}) about the c.m. of the three generated quarks. This statement is correct; so a few of us got together and found 3 new methods that preserve both the original proton c.m. and the correct charge distribution about this c.m.~\cite{CQ2016}. I discuss two of them here along with NQP calculations using the PHENIX2014~\cite{PXPRC89} data.
\subsubsection{Planar Polygon}
Generate one quark at $(r,0,0)$ with $r$ drawn from $r^2 e^{-4.27r}$. Then instead of generating $\cos\theta$ and $\phi$ at random and repeating for the two other quarks as was done by PHENIX2014~\cite{PXPRC89}, imagine that this quark lies on a ring of radius $r$ from the origin and  place the two other quarks on the ring at angles spaced by $2\pi/3$ radians.  Then randomize the orientation of the 3-quark ring spherically symmetric about the origin.  This guarantees that the radial density distribution is correct about the origin and the center of mass of the three quarks is at the origin but leaves the three-quark-triplet on each trial forming an equilateral triangle on the plane of the ring which passes through the origin.   
\subsubsection{Empirical radial distribution, recentered}
\begin{figure}[!hbt] 
      \centering
\raisebox{0.0pc}{\includegraphics[width=0.49\linewidth]{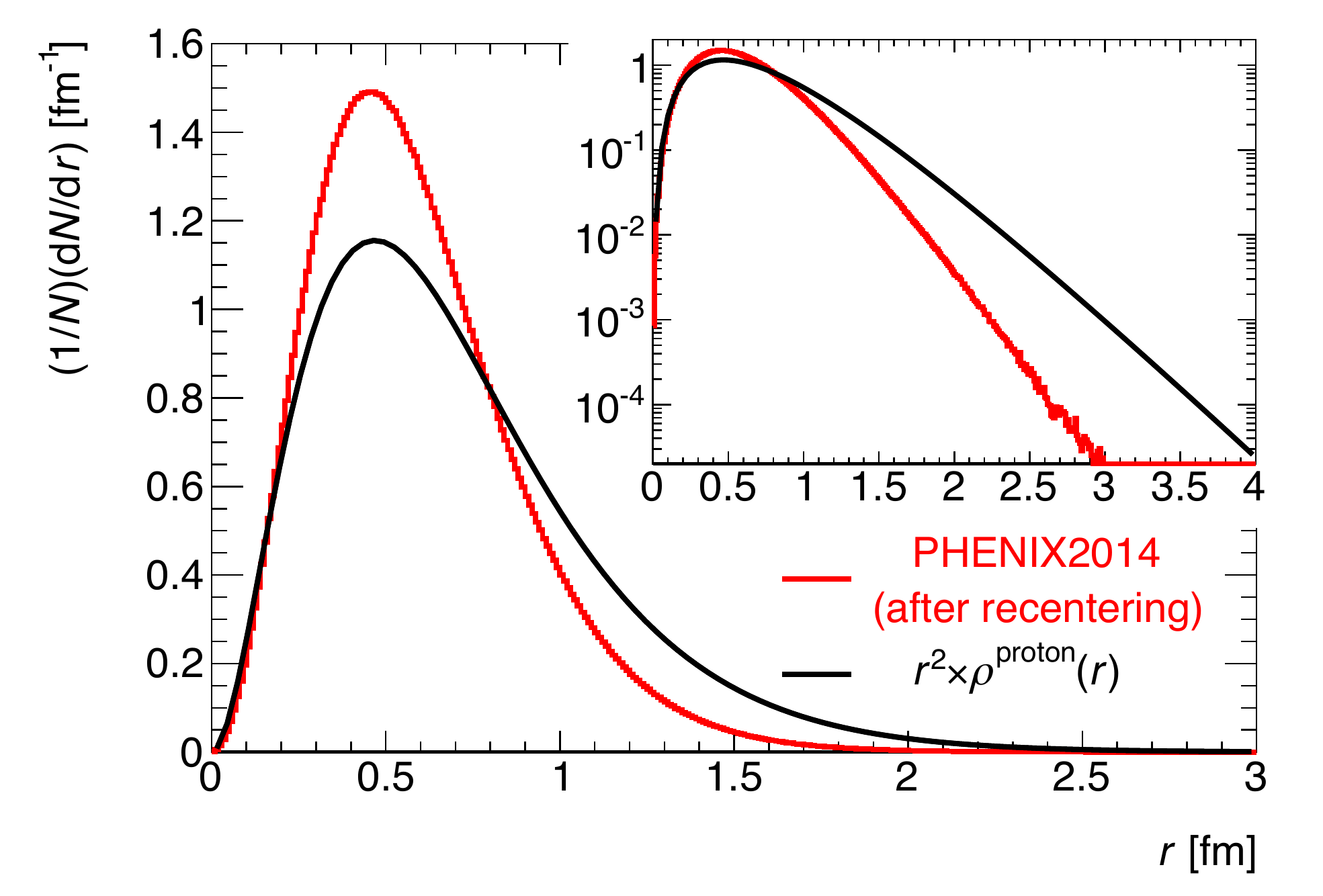}}
\raisebox{0.0pc}{\includegraphics[width=0.43\linewidth]{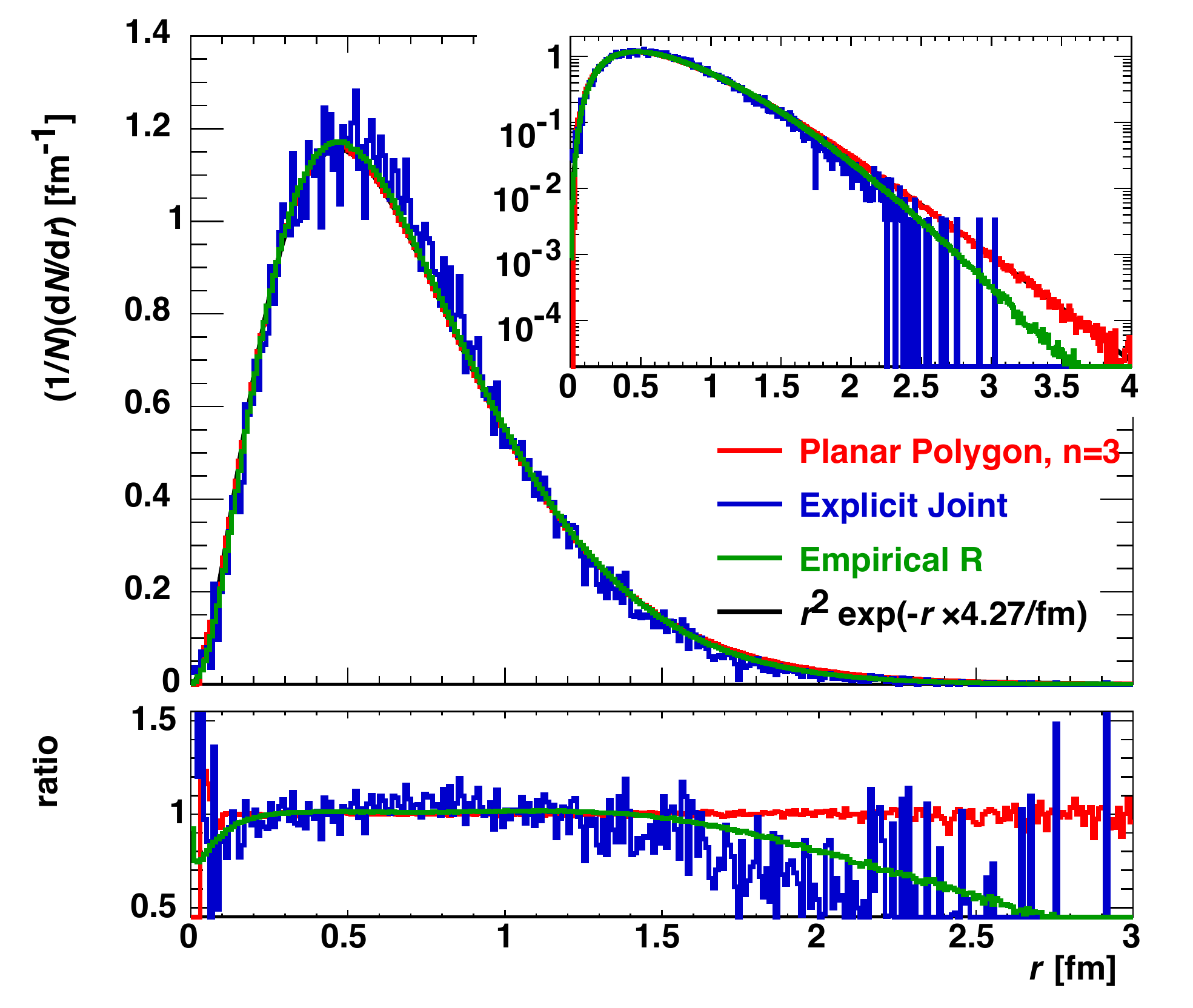}}\vspace*{-1.0pc} 
      \caption[]{\footnotesize a) (left) Radial distribution $d{\cal P}/dr$ about the c.m. of the generated quark-triplets  as a function of $r$ [fm] for the PHENIX2014~\cite{PXPRC89} method compared to $r^2\rho^{\rm proton}(r)$ from Eq.~\ref{eq:Hofstadterdipole} with semi-log plot as inset. b) (right) same for the 3 new methods and the ratios as indicated~\cite{CQ2016}. }
      \label{fig:radial}
   \end{figure}

The three constituent-quark positions are drawn independently from an auxiliary function $f(r)$:
\begin{equation}
   f(r)= r^2 \rho^{\rm proton}(r)\; (1.21466-1.888r+2.03r^{2})\;(1+1.0/r - 0.03/r^{2})\; (1+0.15r) \ .
\label{eq:empirical}
\end{equation}
Then the center of mass of the generated three-quark system is re-centered to the original nucleon position. This function was derived through an iterative, empirical approach. For a given test function $f^{\rm test}(r)$, the resulting radial distribution $\rho^{\rm test}(r)$ was compared to the desired distribution $\rho^{\rm proton}(r)$ in Eq.~\ref{eq:Hofstadterdipole}. The ratio of $\rho^{\rm test}(r) / \rho^{\rm proton}(r)$ was parameterized with a polynomial function of $r$ or $1/r$, and the test function was updated by multiplying it with this parametrization of the ratio. Then, the procedure was repeated with the updated test function $f^{\rm test}(r)$ used to generate an updated $\rho^{\rm test}(r)$ 
until the ratio $\rho^{\rm test}(r) / \rho^{\rm proton}(r)$ was sufficiently close to unity over a wide range of $r$ values. Figure~\ref{fig:radial}~\cite{CQ2016} shows the generated radial distributions compared to $r^2\rho^{\rm proton}(r)$ from Eq.\ref{eq:Hofstadterdipole}.
\subsection{New NQP results using PHENIX2014 data}   
\begin{figure}[hbt] 
      \centering
 \includegraphics[width=0.48\linewidth]{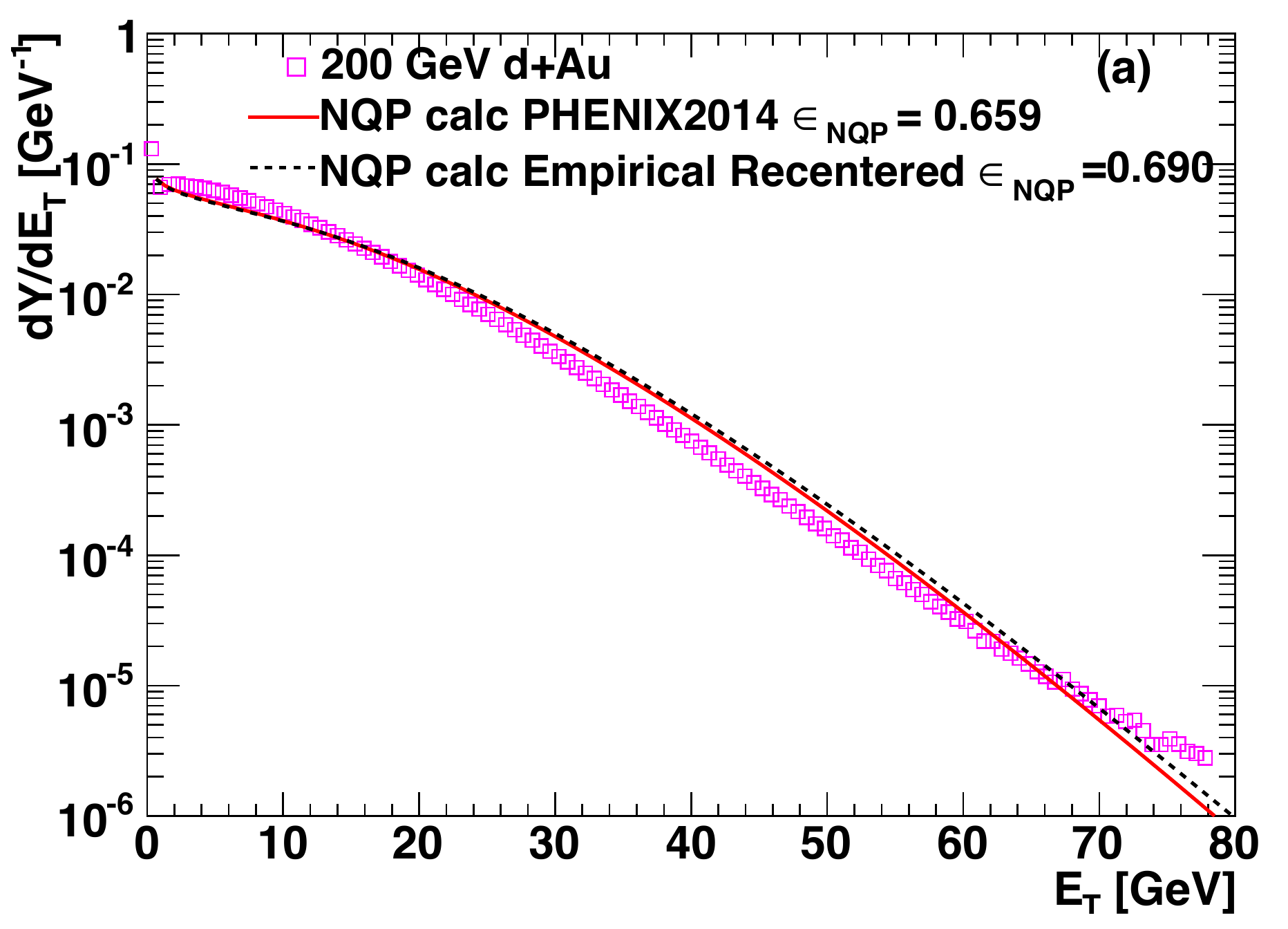}
      \includegraphics[width=0.48\linewidth]{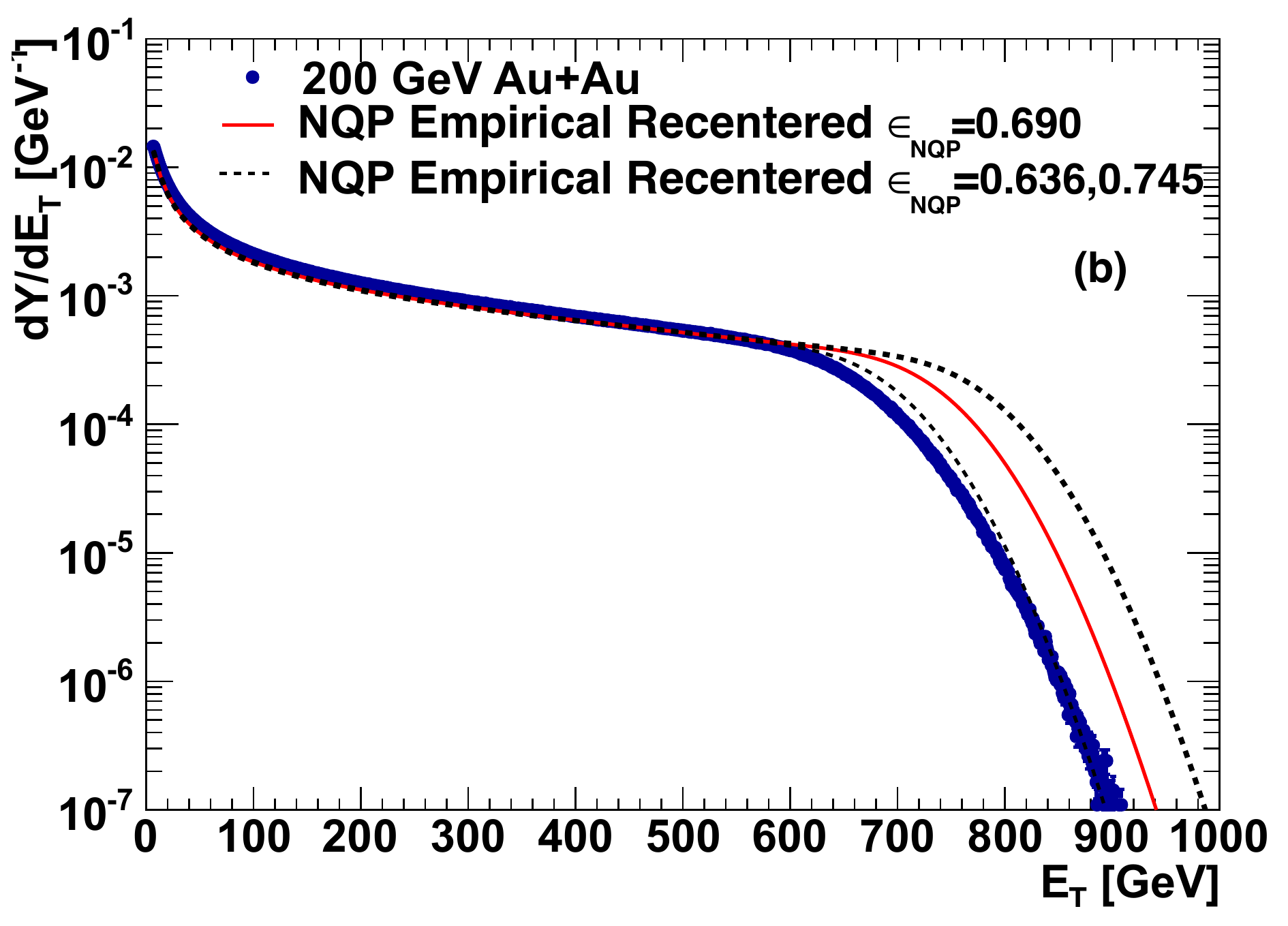} 
      \caption[]{\footnotesize New NQP results~\cite{CQ2016} for $\Et\equiv d\Et/d\eta|_{y=0}$ distributions at $\sqsn=200$ GeV using and compared to PHENIX2014~\cite{PXPRC89} data. a) d$+$Au, b) Au$+$Au.}
\label{fig:newNQP}
   \end{figure}

From Fig.~\ref{fig:radial}b, the Planar Polygon method is identical to Eq.~\ref{eq:Hofstadterdipole} but has all three quarks at the same radius from the c.m. of the proton, which can be tested with more information about constituent-quark correlations in a nucleon. The Empirical recentered method follows $r^2\rho^{\rm proton}(r)$ well out to nearly $r=2$ fm, $Q^2=0.25$ fm$^{-2}=0.01$ GeV$^2$ (compare Fig.~\ref{fig:Mainz}a,b), and is now adopted as the standard. The results of the NQP calculations with the Empirical recentered method~\cite{CQ2016} for the PHENIX2014 data (Fig.~\ref{fig:newNQP}), are in excellent agreement  with the d$+$Au data and agree with the Au$+$Au measurement to within $1 \sigma$ of the calculation (7\% higher in \Et).     
The PHENIX2014 calculation (Fig.~\ref{fig:PXppg100}b) is only $1.2\sigma$ in \Et below the new calculation so that the PHENIX2014 NQP results and conclusions~\cite{PXPRC89} are consistent with the new standard method~\cite{CQ2016}. 
   \vspace*{-2.0pc}
\section{High \pt physics}\vspace*{-1.0pc}
   \begin{figure}[!hbt] 
      \centering
 \includegraphics[width=0.58\linewidth]{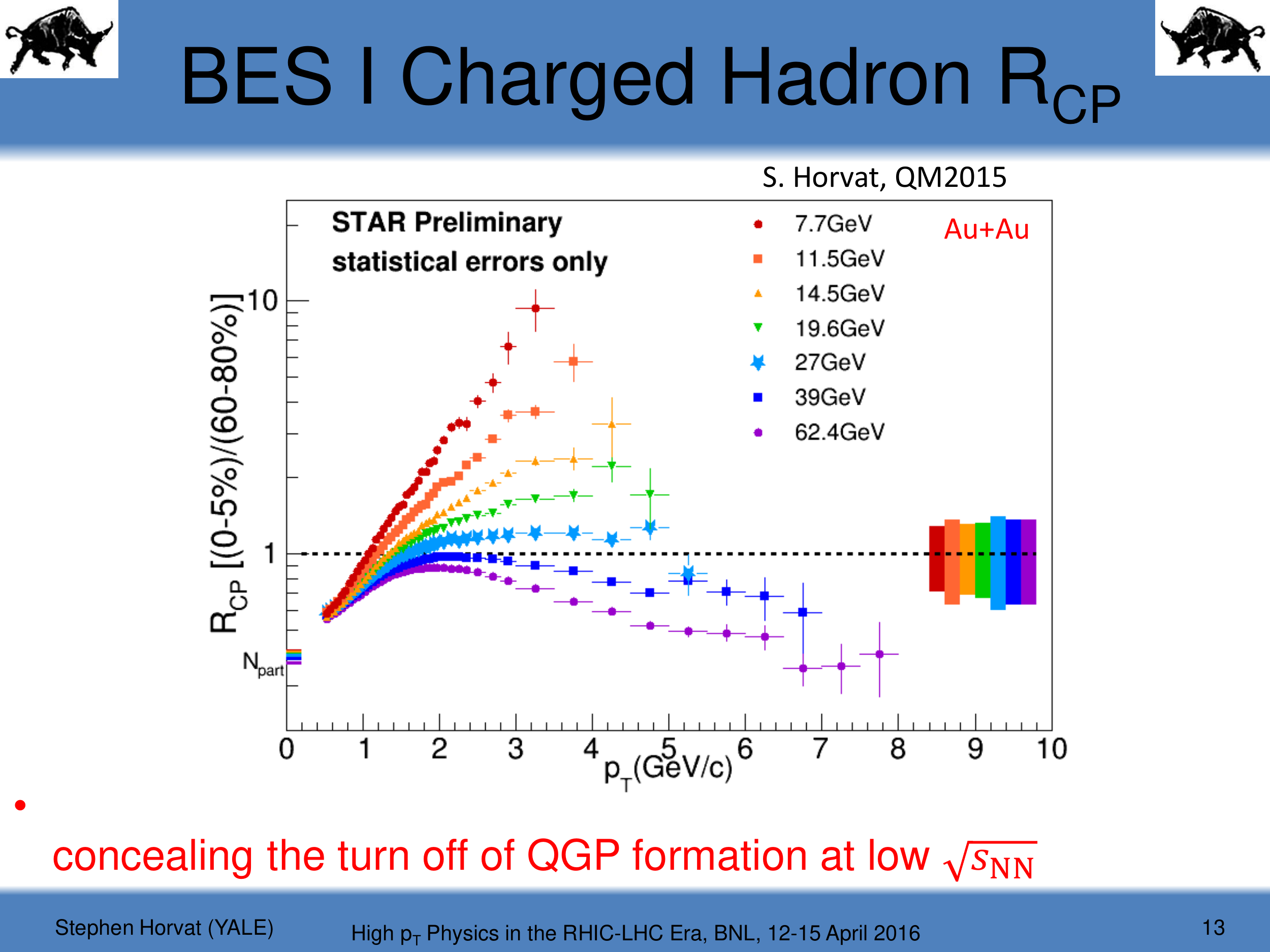}
      \includegraphics[width=0.38\linewidth]{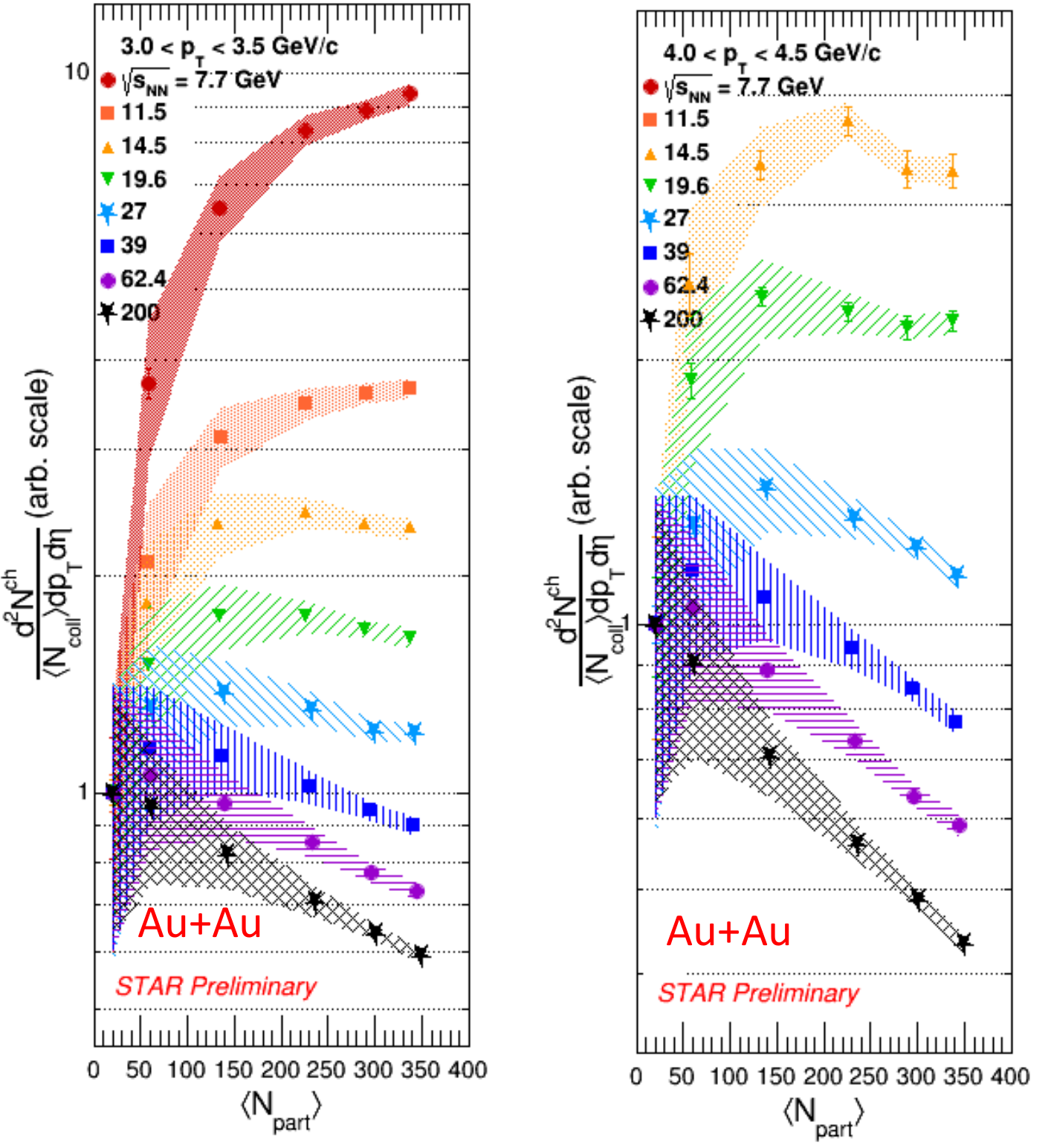}\vspace*{-1.0pc} 
      \caption[]{\footnotesize a)(left) $R_{CP}$ of charged particles vs \sqsn. b) (right) \Ncoll scaling test, $\pt\!\sim$3.2,4.2 GeV/c~\cite{HorvatQM15}}
\label{fig:Horvat}
   \end{figure}
\noindent The suppression of high \pt particles in A$+$A collisons compared to \Ncoll scaled p$+$p measurements is the best evidence for production of the \QGP\ at RHIC. This is presented as defined 
in Eq.~\ref{eq:RAA}  where ${C}$ and ${P}$ represent central and peripheral collisions. 
   \begin{equation}
R_{AA}(p_T)=\frac{(1/N_{AA})\;{d^2N_{AA}/dp_T dy}} {\mean{\Ncoll}\;(1/N_{pp}) {d^2N_{pp}/dp_T dy}}\; {\mbox{or}}\; R_{CP}(p_T)=\frac{\mean{\Ncoll^{P}}(1/N^{C}_{AA})\;{d^2N^{C}_{AA}/dp_T dy}} {\mean{\Ncoll^{C}}\;(1/N^{P}_{AA}) {d^2N^{P}_{AA}/dp_T dy}}
  \label{eq:RAA}\vspace*{-0.0pc} 
 \end{equation}
As indicated on Fig.~\ref{fig:Horvat} there is suppression of $R_{CP}$ for $\sqsn \geq 39$ GeV and no suppression for $\sqsn\leq 27$ GeV. Similarly, a new method, \Ncoll scaling of $R_{AA}(\pt)$ (with the p+p measurement removed)  shows no suppression for $\sqsn\leq 27$ GeV.  Hard-scattering at $\pt\sim4$ GeV/c exists in p$+$p collisions down to \sqs=19.4 GeV~\cite{Cronin79}. Thus, the absence of suppression in Au$+$Au for \mbox{$\sqsn\leq 27$ GeV} suggests the absence of  the \QGP; although this interpretation is complicated by the enhancement in p$+$A observed~\cite{Cronin79} in this same \sqs range.  

\begin{figure}[!hbt] 
      \centering
\includegraphics[width=0.38\linewidth]{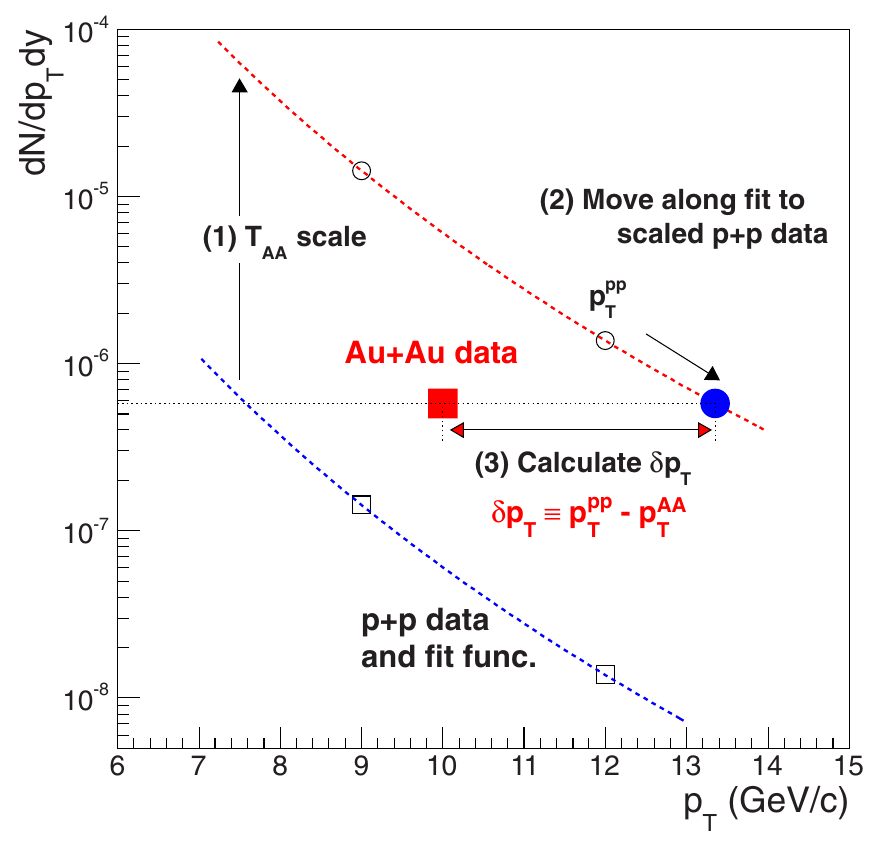}
\includegraphics[width=0.58\linewidth]{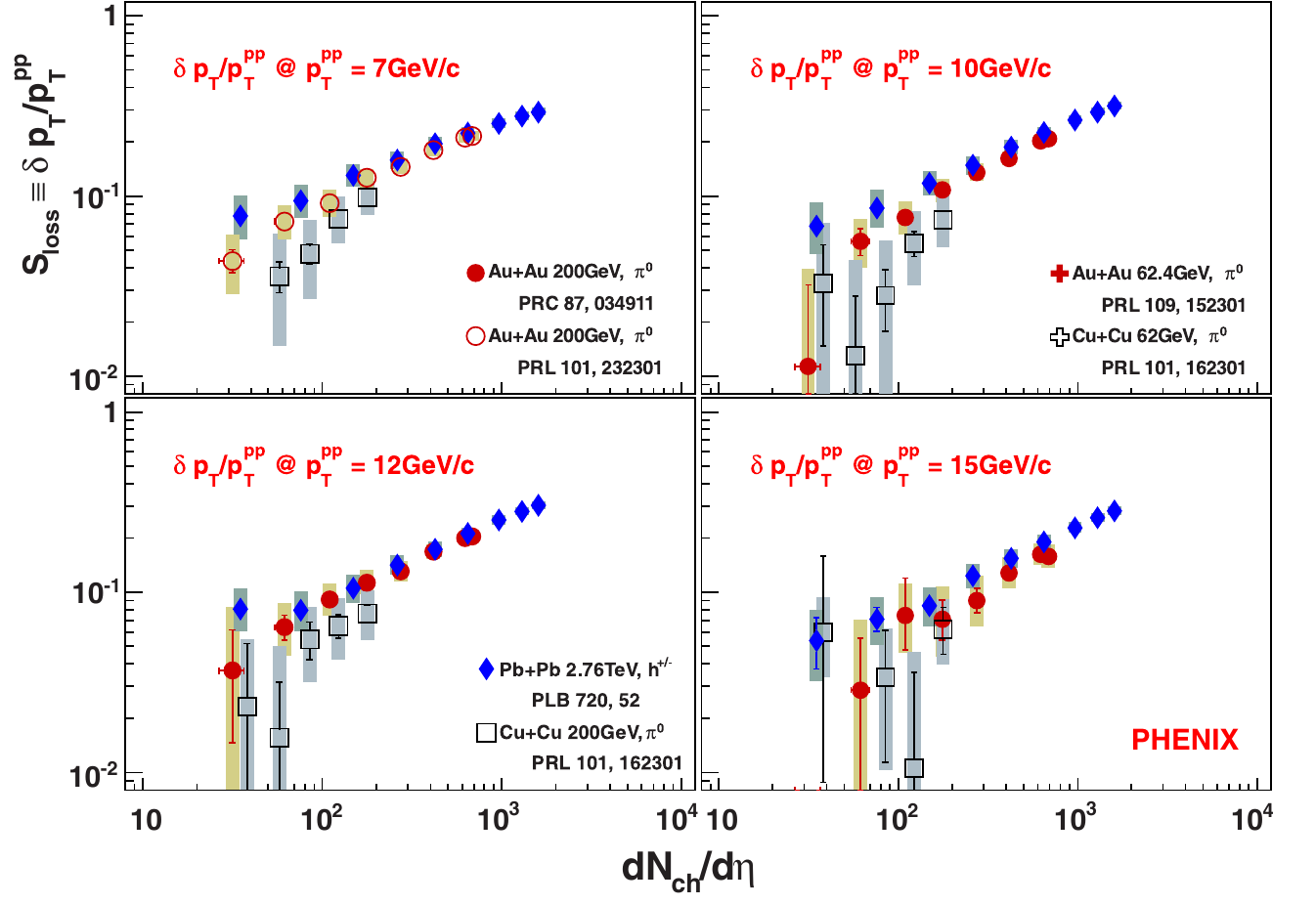} \vspace*{-1.0pc} 
      \caption[]{\footnotesize (left) Schematic of S$_{\rm loss}=\delta \pt/\pt$. b) (right) S$_{\rm loss}$ vs. $d\Nch/d\eta$ as a function of \pt~\cite{PXshiftPRC93}.}
\label{fig:sloss}\vspace*{-2.0pc}
   \end{figure}

    For larger values of \sqsn, especially for comparing LHC to RHIC data, the fractional shift, S$_{\rm loss}=\delta \pt/\pt$, in the $p_T$ spectrum in A$+$A from the expected \Ncoll times the p$+$p value at a given $p_T$ (Fig.~\ref{fig:sloss}) has become more popular than $R_{AA}$~\cite{PXshiftPRC93}.
The data at \mbox{\sqsn=200 GeV} and 2.76 TeV, for $7\leq \pt\leq 15$ GeV/c show a common scaling of S$_{\rm loss}$ with $d\Nch/d\eta$, suggesting that the mid-rapidity multiplicity density may be the key variable for \QGP\ formation. \vspace*{-1pc}  
\subsection{Two-particle azimuthal correlations and $p_{\rm out}$}
An important issue for parton energy loss in a \QGP\ is the broadening of di-jet and di-hadron azimuthal correlations leading to acoplanarity with the beam axis. This effect gives an out-of-plane transverse momentum, $p_{out}$ to a di-jet which is similar to the effect of intrinsic transverse momentum $k_T$ of a parton within a nucleon~\cite{FFFNPB128}. In fact, a theoretical framework (TMD) of parton transverse momentum dynamics within a nucleon, which is not given by perturbative \QCD, has been developed. An early prediction~\cite{CSS85} is that any momentum width sensitive to the nonperturbative intrinsic $k_T$ would grow as the hard-scale (e.g. \pt) increases. 
   \begin{figure}[!ht] 
      \centering
\raisebox{0.0pc}{\includegraphics[width=0.60\linewidth]{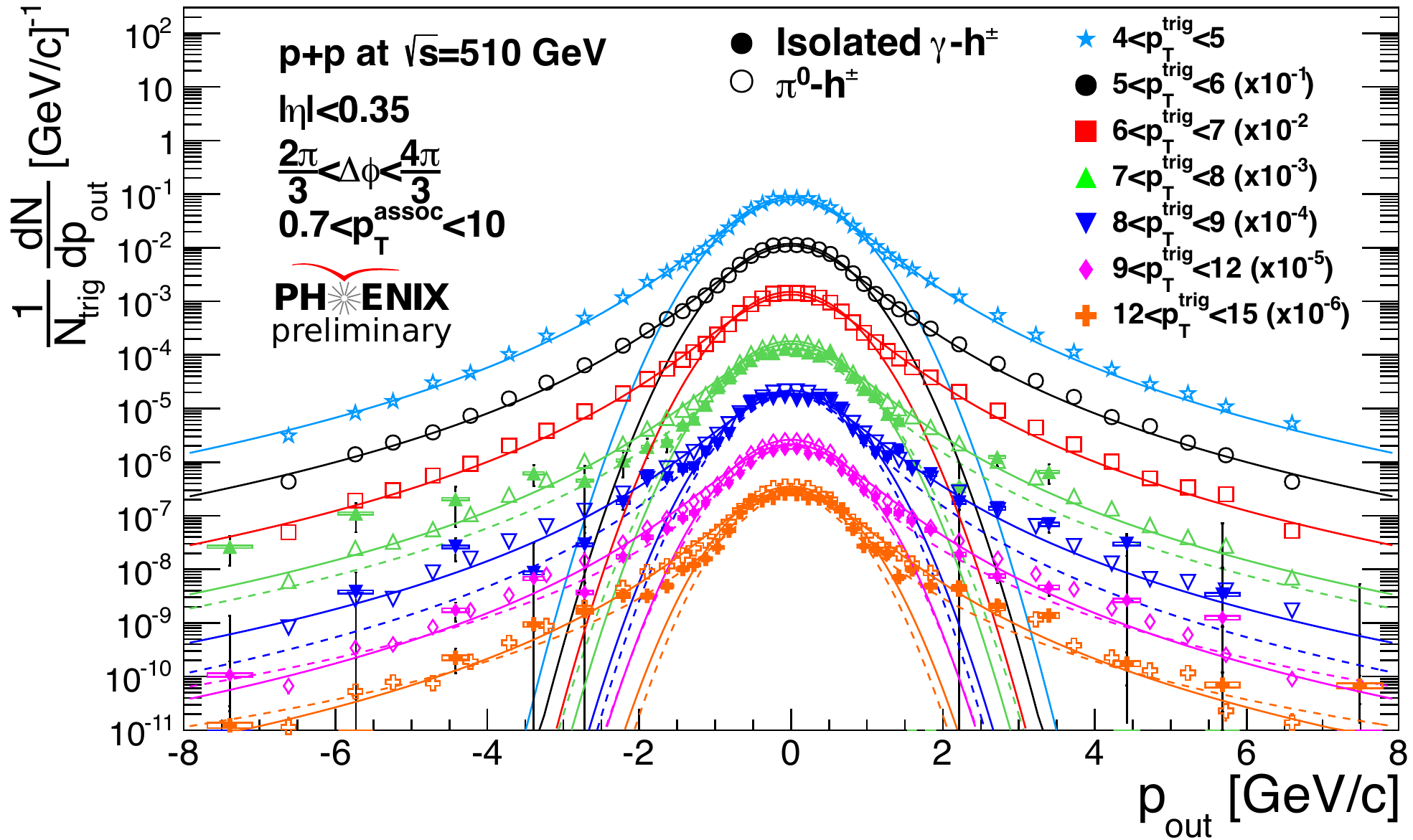}}\hspace*{0.5pc}
\raisebox{1.0pc}{\includegraphics[width=0.38\linewidth]{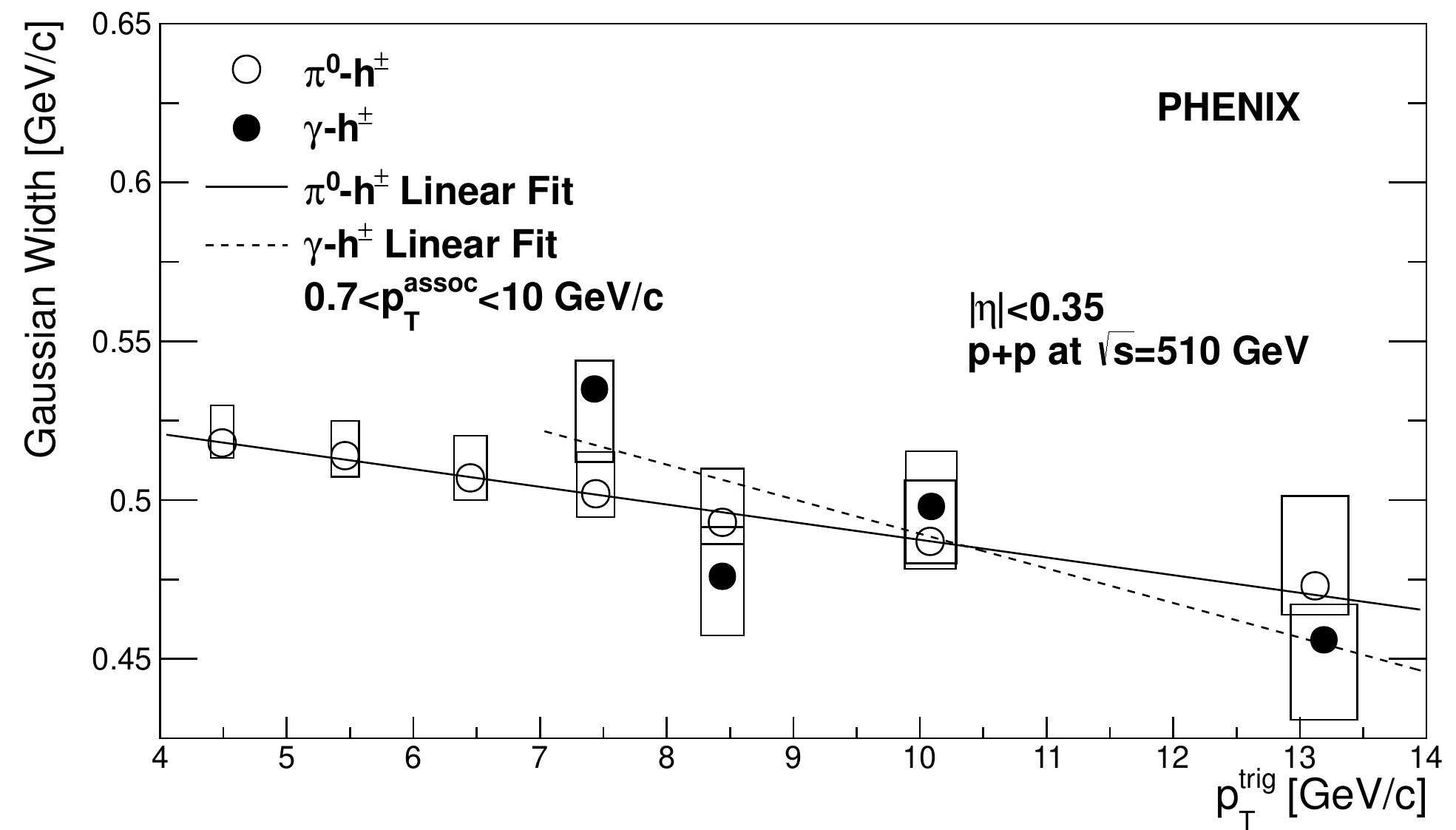}} \vspace*{-1.0pc} 
      \caption[]{\footnotesize a) (left) $p_{out}$ distributions of charged hadrons in $\pi^0+h$ and $\gamma+h$ correlations as a function of $p_{T}^{\rm trig}$ for $0.7<p_{T}^{\rm assoc}<10$ GeV/c~\cite{PXpout}. b) widths of the gaussian $p_{out}$ distributions in (a) vs $p_{T}^{\rm trig}$.  }
\label{fig:pout}
   \end{figure}
A new measurement of  $p_{out}$ from $\pi^0+h$ and $\gamma+h$ correlations in p$+$p at \sqs=510 GeV by PHENIX~\cite{PXpout} (Fig.~\ref{fig:pout}) clearly shows a gaussian distribution for  $p_{out}\lsim 1$ GeV/c, which represents the nonperturbative $k_T$, as well as a perturbative power law tail from gluon emission. The gaussian width as a function of the trigger $p_{T}^{\rm trig}$ shows a decrease with increasing hard scale $p_{T}^{\rm trig}$ that is different from the predicted increase with hard-scale~\cite{CSS85}, clearly indicating the need for a substantial review of the TMD framework.

\end{document}